\documentclass{emulateapj}

\pdfoutput=1
\usepackage{epstopdf}
\usepackage{amsmath}
\usepackage{comment} 
\usepackage{appendix}
\usepackage{enumerate}
\usepackage{natbib}

\usepackage{color}

\newcommand{\about}{$\sim\!\!$~}

\slugcomment{\today}
\newcommand{\xScale}{1.1}   

\begin{document}

\newcommand{\gps}{\ensuremath{g_{\rm P1}}}
\newcommand{\rps}{\ensuremath{r_{\rm P1}}}
\newcommand{\ips}{\ensuremath{i_{\rm P1}}}
\newcommand{\zps}{\ensuremath{z_{\rm P1}}}
\newcommand{\yps}{\ensuremath{y_{\rm P1}}}
\newcommand{\wps}{\ensuremath{w_{\rm P1}}}
\newcommand{\grizy}{\gps\rps\ips\zps\yps}
\newcommand{\griz}{\gps\rps\ips\zps}
\newcommand{\BVRI}{\protect\hbox{$BV\!RI$} }
\newcommand{\UBVRI}{\protect\hbox{$U\!BV\!RI$} }

\newcommand{\PS}{\protect \hbox {Pan-STARRS1}}

\title{Supercal: Cross-Calibration of Multiple Photometric Systems to Improve Cosmological Measurements with Type  I\lowercase{a} Supernovae}

\shorttitle{Cross-calibration of Multiple Photometric Systems}
\shortauthors{Scolnic et al.}
\email{Electronic Address: dscolnic@kicp.uchicago.edu}

\def\uch{1}
\def\stsci{2}
\def\jhu{3}
\def\mpia{4}
\def\urba{5}
\def\urbp{6}
\def\harvard{7}
\def\gmto{8}
\def\hawaii{9}
\def\durham{10}
\def\princeton{11}
\def\naval{12}
\def\CfA{13}

\author{
D. Scolnic\altaffilmark{\uch},
S. Casertano\altaffilmark{\stsci},
A. Riess\altaffilmark{\stsci,\jhu},
A. Rest\altaffilmark{\stsci},
E. Schlafly\altaffilmark{\mpia},
R.~J. Foley\altaffilmark{\urba,\urbp},
D. Finkbeiner\altaffilmark{\harvard},
C. Tang\altaffilmark{\urba},
W.~S. Burgett\altaffilmark{\gmto},
K.~C. Chambers\altaffilmark{\hawaii},
P.~W. Draper\altaffilmark{\durham},
H.~Flewelling\altaffilmark{\hawaii},
K.~W. Hodapp\altaffilmark{\hawaii},
M.~E. Huber\altaffilmark{\hawaii},
N. Kaiser\altaffilmark{\hawaii},
R.~P. Kudritzki\altaffilmark{\hawaii},
E.~A. Magnier\altaffilmark{\hawaii},
N. Metcalfe\altaffilmark{\durham},
C.~W. Stubbs\altaffilmark{\harvard,\CfA}
}

\altaffiltext{\uch}{Department of Physics, The University of Chicago, Chicago,IL 60637, USA}
\altaffiltext{\stsci}{Space Telescope Science Institute, 3700 San Martin Drive, Baltimore, MD 21218, USA}
\altaffiltext{\jhu}{Department of Physics and Astronomy, Johns Hopkins University, 3400 North Charles Street, Baltimore, MD 21218, USA}
\altaffiltext{\mpia}{Max Planck Institute for Astronomy, K\:onigstuhl 17, D-69117 Heidelberg, Germany}
\altaffiltext{\urba}{Astronomy Department, University of Illinois at Urbana-Champaign, 1002 West Green Street, Urbana, IL 61801, USA}
\altaffiltext{\urbp}{Department of Physics, University of Illinois Urbana-Champaign, 1110 West Green Street, Urbana, IL 61801, USA}
\altaffiltext{\harvard}{Department of Physics, Harvard University, 17 Oxford Street, Cambridge MA 02138}
\altaffiltext{\gmto}{GMTO Corporation, 251 S. Lake Ave., Suite 300, Pasadena, CA 91101 USA}
\altaffiltext{\hawaii}{Institute for Astronomy, University of Hawaii, 2680 Woodlawn Drive, Honolulu, HI 96822, USA}

\altaffiltext{\durham}{Department of Physics, University of Durham Science Laboratories, South Road Durham DH1 3LE, UK}
\altaffiltext{\princeton}{Department of Astrophysical Sciences, Princeton University, Princeton, NJ 08544, USA}
\altaffiltext{\naval}{US Naval Observatory, Flagstaff Station, Flagstaff, AZ 86001, USA}
\altaffiltext{\CfA}{Harvard-Smithsonian Center for Astrophysics, 60 Garden Street, Cambridge, MA 02138, USA}

\begin{abstract}
Current cosmological analyses which use Type Ia supernova (SN\,Ia) observations combine SN samples to expand the redshift range beyond that of a single sample and increase the overall sample size.  The inhomogeneous photometric calibration between different SN samples is one of the largest systematic uncertainties of the cosmological parameter estimation.  To place these different samples on a single system, analyses currently use observations of a small sample of very bright flux standards on the \textit{HST} system.  We propose a complementary method, called `Supercal', in which we use measurements of secondary standards in each system, compare these to measurements of the same stars in the Pan-STARRS1 (PS1) system, and determine offsets for each system relative to PS1, placing all SN observations on a single, consistent photometric system.  PS1 has observed $3\pi$ of the sky and has a relative calibration of better than 5 mmag (for $\sim15<griz<21$ mag), making it an ideal reference system.  We use this process to recalibrate optical observations taken by the following SN samples: PS1, SNLS, SDSS, CSP, and CfA1-4.  We measure discrepancies on average of 10 mmag, but up to 35 mmag, in various optical passbands.  We find that correcting for these differences changes recovered values for the dark energy equation-of-state parameter, $w$, by on average $2.6\%$.  This change is roughly half the size of current statistical constraints on $w$.  The size of this effect strongly depends on the error in the $B-V$ calibration of the low-$z$ surveys.  The Supercal method will allow future analyses to tie past samples to the best calibrated sample. 
\end{abstract}

\section{Introduction}
\label{sec:intro}

Since the initial discovery of cosmic accelerating expansion (\citealp{Riess98}, \citealp{Perlmutter99}), many samples of Type Ia supernovae (SN\,Ia) have been acquired to better constrain the dark energy equation-of-state parameter, w.  As the systematic uncertainties in joint samples are nearly equal to the statistical uncertainties, increasing effort must be expended on reducing systematic uncertainties so that the total uncertainties do not soon hit a systematic floor.  Of all the systematic uncertainties, recent analyses (e.g., \citealp{Betoule14}, \citealp{Scolnic14b} [hereafter S14]) found that those related to photometric calibration make up $>70\%$ of the total systematic uncertainty and pose the most immediate challenge.

Most SN analyses that attempt to constrain $w$ combine publicly available SN samples to improve statistics and cover a wider redshift range.  To reproduce the most recent cosmological results \citep{Betoule14, Rest14}, one must combine SN from $>$10 independently calibrated photometric systems.  The calibration of each system is performed by multiple groups using sometimes significantly different methodology.   To date, no cosmology analysis with SN\,Ia finds a solution so that the calibration of all the various systems is consistent.   Without doing so, a SN analysis likely underestimates the systematic uncertainties of the cosmological parameters.  \cite{Betoule14} showed that there are differences of $5\%$ between average distances of a single sample relative to the expected distances from the $\Lambda$CDM model.  It is now imperative to determine whether the small deviations are due to noise, calibration uncertainties, or deviations from the $\Lambda$CDM model.

In this analysis, we take advantage of the uniform calibration of the PS1 survey with $<1\%$ precision and accuracy \citep{Schlafly12} in order to measure and improve the consistency of catalog photometry between different surveys.  Here, we use publicly available data from  \textit{HST}, SDSS, SNLS, CSP, and CfA1-4.  In \S\ref{sec:current}, we discuss how each sample is currently calibrated.  We introduce an analysis of the cross-calibration between multiple systems in \S\ref{sec:cross}.   In \S\ref{sec:consequences}, we discuss the magnitude of the discrepancies and the implications of correcting for these, including effects on SNIa distances and recovered cosmology.  We also quantify the dominant uncertainties in this approach.  Our discussion and conclusions are in \S\ref{sec:discuss} and \S\ref{sec:conclus}.

\section{Current Calibration}
\nobreak
\label{sec:current}

In recent cosmological analyses, the calibration of each system may be compared directly or indirectly to the \textrm{AB} system (\citealp{Oke83}, \citealp{Fukugita96}).  In the \textrm{AB} system, a monochromatic magnitude is defined such that
\begin{equation}
m_{\textrm{AB}}(\nu)  = -2.5~\log_{10} \bigg(\frac{f_{\nu}}{1~\rm{Jy}} \bigg) + 8.90~\textrm {mag},
\end{equation}
where $f_{\nu}$ is the flux per unit frequency from an object in Jy.  Therefore, a magnitude 0 object should have the same counts as a source of $f_{\nu}  = 3631$~Jy.

We can define an \textrm{AB} {\em broadband} magnitude by the following
equation:
\begin{equation}
m_{\textrm{AB}} = 2.5 \times \log_{10} \frac{ \int (h\nu)^{-1}p(\nu) f_{\nu}~ d\nu }{ \int (h\nu)^{-1}p(\nu)~3631~\rm{Jy}~\mathit{d\nu}}\label{eq:broadbandAB}
\end{equation}
where $p(\nu)$ is the filter response function and this equation assumes that the detector is a photon-counting device.

Some of the surveys analyzed here calibrate their photometry to the AB system while others use Vega-calibrated systems.  In either case, an \textrm{AB} offset can be given to convert the zeropoint of the calibration of one measured system to the true AB system.  For a given system $S$ and passband $p$, this may be expressed as
\begin{equation}
m^{S,p}_{\rm AB}=m^{{ S,p}}_{\rm Sys}+\Delta^{{S,p}}_{\rm AB}.
\end{equation}
where $m^{{ S,p}}_{\rm Sys}$ is the system magnitude, $m^{S,p}_{\rm AB}$ is the AB magnitude and $\Delta^{{S,p}}_{\rm AB}$ is the offset for a particular filter between the system magnitude and the AB magnitude.

By Eq. 2, these offsets can be found explicitly if given a spectrum $f_{\nu}$ defined to be on the AB system, a measured passband $p_{\nu}$ and the observed magnitude of the star in system $S$.  For all systems below, the spectra used for this process are taken of \textit{HST} Calspec standards \citep{1996AJ....111.1743B}.  These spectra are composite spectra from STIS and NICMOS observations and have an uncertainty of \about5 mmag for every $5000~\rm \AA$ (Bohlin 2014) from $3000~ \rm \AA$ to $15000~ \rm \AA$.

There are two major components to the systematic uncertainties of each system's calibration: how well observations are tied to fundamental photometric standards and how well the photometric standards themselves are calibrated.  The advantage of the calibration method for each survey discussed below is that it relies on observations of known \textit{HST} Calspec standards for which we have accurate spectra and can compare synthetic photometry with observational photometry.  The disadvantage of this approach is that the Calspec standards are sparse (\about20 over the observable sky) and typically much brighter ($r<13$ mag) than normal survey stars ($r>15$ mag).  Furthermore, different surveys use different Calspec standards.

\begin{figure}
\centering
\epsscale{1.15}  
\plotone{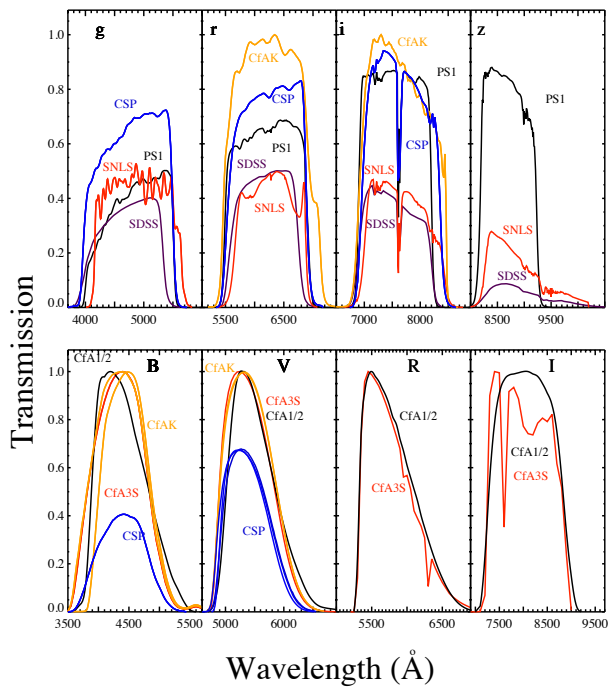}
\caption{The filter transmission functions of all the systems in this analysis.  Comparisons are broken into $griz$ and $\BVRI$.  For both CSP and CfA-Keplercam (CfAK), multiple filters are shown as in different periods of the survey, different filters were used.}
\label{fig:filters}
\end{figure}

Below, we briefly review the calibration of each of the photometric catalogs used in this analysis. We only analyze publicly available sources.  A summary of the filters, calibration standards, and systematic uncertainties of each system is given in Table 1.   The systematic uncertainties for each system are described such that there is an uncertainty in the common zeropoint to observations of a given filter and an uncertainty in the mean wavelength of that filter.  The transmission functions of all the filters from the various systems analyzed are shown in Fig.~\ref{fig:filters}.
 
 This section is separated into an explanation of the calibration of the PS1 photometric system, the systems of the other higher-$z$ surveys and systems of the low-$z$ surveys.  The PS1, SNLS and SDSS systems share a common calibration path in that multiple Calspec standards are used to define the photometry on the AB system and nightly photometry can be tied directly to stellar catalogs in each survey's natural system.  The low-$z$ systems are partly tied to the Vega system and partly tied to the AB system, and only BD+17$^{\circ}4708$ (hereafter BD+17) is used to tie magnitudes from each filter to the AB system.  The AB offsets for all surveys are given in Table 1.  These offsets are the same ones applied when fitting light curves of the SN.
 
 For the low-$z$ surveys, zeropoints of the nightly photometry are determined either by transforming Landolt \citep{Landolt07} and Smith standards  \citep{Smith02} onto the respective natural systems or by transforming the nightly photometry onto the system of Landolt standards.  The systematic uncertainties of the zeropoints given in Table 1 should therefore include: uncertainties in the AB magnitudes of the primary standard(s) used, uncertainties in the measurements of the primary standard(s) by each survey, uncertainties in the transfer of zeropoints between the local standards to the primary standards, and systematic uncertainties in the measurements of the local standards.

\begin{table*}[ht]
\caption{Previously reported calibration differences between systems}
\begin{tabular}{l|ccccccc|}
\hline \hline
System & Filters & Standards & Stan. Observation & ZP Err & Wave Err &AB offsets & Ref. \\
~ & ~ & ~ & ~ & [mmag] & [nm] & [mmag] & ~ \\

\hline
PS1 & $griz$ & 7* & PS1 & [12,12,12,12] & [0.7,0.7,0.7,0.7] & [-23,-33,-24,-24] & T12,S14,S15\\
SNLS & $griz$ & 3* & SNLS & [3,6,4,8] & [0.3,3.7,3.1,0.6]  & [-7,-8,-13,4.7] & B13\\
SDSS & $ugriz$ & 3* & SDSS PT & [8,4,2,3,5] & [0.6,0.6,0.6,0.6,0.6] & [-,-28,-14,-27,-20] & M8,D10,B13\\
CfA1/2 & \UBVRI& BD+17 & Landolt & [100,15,15,15,15] & [2.5,1.2,1.2,2.5,2.5] & [-,131,6,-168,-410] &Land\\
CSP & $ugri.BV$& BD+17 & Smith  Landolt  & [23,9,8,7,8,8] &[0.7,0.8,0.4,0.2,0.7,0.3] & [-,10,-4,-13,102,6/-7/2] & S11\\
CfAK &$ri.UBV$& BD+17 & Smith Landolt  & [25,7,31,11,7] & [0.7,0.7,2.5,0.7,0.7] & [-3,-9,-,80,6] & H09a\\
CfAS & \UBVRI& BD+17 & Landolt & [70,11,7,7,20] & [2.5,0.7,0.7,0.7,0.7] & [-,71,7,-179,-406] & H09a\\
CfA4 & $ri.BV$& BD+17 & Smith Landolt & [31,11,7,25,7]  &  [2.5,0.7,0.7,0.7,0.7] & [-3/1,-9,80/135,6] &H09b\\
\hline
\end{tabular}
Summary of various sytems used in this analysis.  The columns are: \textbf{Filters} used for observations, \textbf{Standards} used to determine the absolute flux zeropoints, \textbf{Standard Observations} that details with what telescope/camera the standards were originally observed with, \textbf{zeropoint error} claimed by the survey, \textbf{wavelength error} of the filter bandpasses, \textbf{AB offsets} to transform from the system magnitudes to AB magnitudes, and \textbf{primary reference}.  The standards used by SNLS and SDSS are G191B2B, GD153, and GD71.  The standards used by PS1 are given in Table 2.  AB offsets for the low-$z$ systems are generally large as the system was not defined on the AB system.  Multiple values for certain offsets indicate multiple periods of a survey where the filters changed.
\end{table*}

\subsection{Calibration of the PS1 full sky sample}

\textbf{PS1} - The PS1 photometric calibration is presented in \cite{Tonry12}, hereafter T12.  PS1 is a 1.8 m telescope on Haleakala with a field of view of 7 square degrees.  The observation strategy has two large parts: a $3\pi$ survey across the entire observable sky and a Medium Deep survey for 10 fields of 7 square degrees each.  Both surveys observe in $\grizy$ (see Figure 1 for $\griz$ filters used for this analysis).  T12, based on work from \cite{Stubbs10} and \cite{Tonry11}, uses an innovative laser diode system to accurately and precisely determine the filter bandpass edges  ($\sigma_{\lambda}<7~\rm \AA$) and throughput curves.   The flux calibration of PS1 measurements relies on an iterative process that includes work from T12 and \cite{Schlafly12} and is augmented by S14.  T12 analyzes observations of 7 \textit{HST} Calspec standards with PS1 and compares the observed magnitudes of these standards to the predicted magnitudes from synthetic photometry.  T12 then finds the AB offsets (Eq. 3) so that the observed magnitudes best matches the synthetic photometry, given fixed constraints from measurements on the bandpass edges and shapes.  

It is necessary that photometry from all observations from the $3\pi$ survey and Medium Deep survey can be linked by a single zeropoint for each filter.  To do so, \cite{Schlafly12} uses the `Ubercal' process \citep{Padmanabhan} that creates a relative calibration across the sky and directly ties the zeropoints of all photometry to the catalogs from the fields that contain the observed Calspec standards analyzed in T12.   Because PS1 repeatedly observed the same regions \about 12 times in each filter and there are overlapping regions between different pointings, all relative zeropoints can be determined simultaneously and robustly.  This procedure determines the system throughput, atmospheric transparency, and large-scale detector flat field.  The solution from this process includes data from both the Medium Deep fields and the $3\pi$ sky over the full survey.  From Ubercal, new photometric catalogs are created across the entire observable sky with relative accuracy and precision better than 5 mmag.  To improve on the initial zeropoints given by T12, S14 then iterates on this process by analyzing the entire sample of Calspec standards observed over the course of the $3\pi$ survey and redetermines the AB offsets for the PS1 system.  S14 also analyzes the variation of filter transmission functions across the focal plane and finds the variations across the focal plane with a radial dependence for stars with $0.4<\gps-\ips<1.5$ to be up to $8$ mmag (due to the design of the filters), though typically at a dispersion level of \about2 mmag.  The filter throughput is measured at multiple radial positions, and brightnesses can be corrected based on radial position on the focal plane.  
\begin{table}[ht]
\centering
\caption{PS1 Observed and Synthetic Magnitudes of Calspec standards}
\begin{tabular}{lllllllllllll}
\hline \hline
Star & Filter & Obs. Magnitude & Syn. Magnitude \\
\hline
SF1615 & \gps & 16.975 (0.007) & 16.996\\
Snap-2 & \gps & 16.413 (0.008) & 16.447\\
Snap-1 & \gps & 15.477 (0.007) & 15.495\\
WD1657+343 & \gps & 16.212 (0.007) & 16.224\\
KF06T2 & \gps & 14.391 (0.007) & 14.429\\
lds749b & \gps & 14.562 (0.010) & 14.573\\
C26202 & \gps & 16.651 (0.008) & 16.676\\
SF1615 & \rps & 16.523 (0.007) & 16.562\\
Snap-2 & \rps & 16.012 (0.008) & 16.046\\
Snap-1 & \rps & 15.857 (0.007) & 15.894\\
WD1657+343 & \rps & 16.669 (0.007) & 16.693\\
KF06T2 & \rps & 13.573 (0.010) & 13.606\\
lds749b & \rps & 14.765 (0.008) & 14.808\\
C26202 & \rps & 16.347 (0.007) & 16.368\\
SF1615 & \ips & 16.360 (0.006) & 16.385\\
Snap-2 &  \ips  & 15.878 (0.007) & 15.904\\
Snap-1 &  \ips  & 16.191 (0.007) & 16.202\\
WD1657+343 &  \ips  & 17.053 (0.006) & 17.073\\
lds749b &  \ips  & 15.000 (0.010) & 15.039\\
C26202 &  \ips  & 16.240 (0.008) & 16.263\\
GD153 & \zps & 14.230 (0.007) & 14.263\\
P177D & \zps & 13.137 (0.008) & 13.154\\
SF1615 & \zps & 16.285 (0.006) & 16.318\\
Snap-2 & \zps & 15.846 (0.006) & 15.875\\
Snap-1 & \zps & 16.393 (0.006) & 16.424\\
WD1657+343 & \zps & 17.346 (0.007) & 17.360\\
KF06T2 & \zps & 13.083 (0.010) & 13.084\\
C26202 & \zps & 16.211 (0.006) & 16.245\\
GD153 & \yps & 14.450 (0.007) &  14.472\\
P177D & \yps & 13.135 (0.007) &  13.135\\
SF1615 & \yps & 16.269 (0.006) &  16.278\\
Snap-2 & \yps & 15.837 (0.007) &  15.853\\
Snap-1 & \yps & 16.581 (0.007) &  16.567\\
WD1657+343 & \yps & 17.570 (0.006) &  17.579\\
KF06T2 & \yps & 12.980 (0.007) &  12.991\\
lds749b & \yps & 15.380 (0.008) &  15.401\\
C26202 & \yps & 16.225 (0.007) &  16.251\\\\
\hline 
\end{tabular}
~~~~~~~~~~~~~~~~~\\The calspec standard stars with adequate PS1 photometry are presented here.  Both the observed and synthetic magnitudes of these standards are given, before an AB correction is applied to the natural PS1 magnitudes (from Tonry et al. 2012). The uncertainties in these measurements are given in brackets.  The synthetic spectra can be found on the Calspec website\footnote{http://www.stsci.edu/hst/observatory/crds/calspec.html}; we use version 005 in this analysis. 
\end{table}

For the present analysis, we repeat the process done in S14 to redetermine the AB offsets for the PS1 system.  We find that the majority of the observations of the Calspec standards from T12 placed the standards at the same position on the same chip for each observation.  This position was very close to the center of the focal plane where it has been noted that there is a strong gradient in the behavior of the chip \citep{Rest14}.  The Ubercal solution for the large-scale detector flat field \citep{Schlafly12} at the location where these standards were observed varies by up to $20$ mmag over less than a quarter of a CCD.  Given this issue, observations from T12 are not included in the present analysis.  Also, only observations fainter than the saturation limit of  [14.3,~14.4,~14.6,~14.1]  in \gps\rps\ips\zps\ are included.  Table 2 shows both the synthetic photometry of the six Calspec standards and the Ubercal photometry of these standards.  As shown in Table 3, we find that corrections from S14 of 20-35 mmag in each filter are needed.  The significant size of these corrections is partly due to the update of the \textit{HST} Calspec spectra.  Further information will be given in the public release of the PS1 data.

\begin{table}[ht]
\caption{PS1 Photometric System}
  \begin{center}
    \begin{tabular}{l | c c | c c}
    Filter & S14 & S15 & S14-S15 & S14-S15 \\
    ~  & [mmag] & [mmag] & ~ & ~ \\
    ~  & ~ & ~ & HST-Recal & Improved Phot \\

    \hline
\gps &  $-8 \pm 12$ & $-20 \pm 8.0$ & $-7$ & $-13$\\
 \rps & $-10 \pm 12$ & $-33 \pm 8.0$ & $-9$ & $-24$\\
\ips &    $-4 \pm 12$ & $-24 \pm 8.0$ & $-9$ & $-15$\\
     \zps &    $-7 \pm 12$ & $-28 \pm 8.0$ & $-9$  & $-12$ \\
    \hline
    \end{tabular}
  \end{center}
  Corrections of the AB offsets from the original definition of the PS1 calibration as given in in T12.  The S15 include the updates to the most current \textit{HST} Calspec magnitudes (version 005).  The breakdown of the change in values due to the HST version update and from our own improved photometry is given in the last two columns respectively.  The list of Calspec standards and both their synthetic and observed magnitudes used for this recalibration is given in Table 2.
\end{table}

\subsection{Calibration of intermediate and high-$z$ surveys}

\textbf{ SDSS-II } - The basis for the Sloan Digital Sky Survey-II (hereafter referred to as `SDSS') calibration is presented in \cite{Holtzman08} and \cite{Doi10}, with important updates given in \cite{Betoule12}.  The primary instrument of the SDSS Supernova Survey is the SDSS CCD camera \citep{Gunn98} which was mounted on a dedicated 2.5 m telescope \citep{Gunn06} at Apache Point Observatory (APO), New Mexico. The survey observed in the five optical bands: $ugriz$ (\citealp{Fukugita96}, see Figure 1 for $griz$ filters used for this analysis).  The survey covers a 300 square-degree region ($2.5^{\circ}$ wide over 8 hours in right ascension). The photometry explored in this analysis is of the region centered on the celestial equator referred to as `Stripe 82'.  The measurement of SDSS effective passbands are described in \cite{Doi10}.

The absolute flux calibration has been determined using the SDSS Photometric Telescope (PT) observations of Calspec solar analog stars \citep{Tucker06}. This process is described in detail in \cite{Holtzman08}. \cite{Betoule12} updated the AB offsets to reflect recent revisions to the \textit{HST} Calspec observations and SDSS observations. Small differences between the individual filters mounted on the different columns of the camera can be neglected due to the SDSS calibration strategy \citep{Betoule12}.  

\textbf{SNLS } - The latest calibration of the Supernova Legacy Survey (SNLS) is given in \cite{Betoule12}. SNLS uses the 3.6 m CFHT atop Mauna Kea.  SNLS covers the four low extinction fields of the CFHT Legacy Survey Deep component (called D1 to D4).  The fields are repeatedly imaged in the 4 optical bands $g_M$,$r_M$,$i_M$,$z_M$ (see Figure 1).  The original $i_M$ filter was broken in July 2007 and replaced by a slightly different $i_M$ (denoted $i2_M$) filter in October of the same year.  As $i_M$ magnitudes are given for all SN observations published, and not $i2_M$, we exclude $i2_M$ from this analysis.  The field-of-view of the MegaCam camera is $0.96 \times 0.94$ deg$^2$ \citep{Boulade03}.  Measurements of the SNLS bandpasses are presented in \cite{Regnault09}.

The SNLS calibration is determined from multiple paths including observations of \textit{HST} Calspec standards and Landolt stars.  Spatial variations of the passband response result in variation in the brightness found for stars when they are observed at different positions on the MegaCam focal plane (R09).  In the SNLS data release, magnitudes of the stars are transformed as if they were observed at the center of the focal plane.

\subsection{Calibration of low-$z$ surveys}

In recent SN analyses (\citealp{Betoule14}, S14), the calibration of the absolute flux of the various low-$z$ surveys is tied to measurements of the primary standard star  BD+17.  For filters $ugri$, the flux is calibrated to the \cite{Smith02} magnitudes of BD+17.  For filters \UBVRI, the flux is calibrated to the Landolt magnitudes of BD+17.  Magnitudes from \cite{Smith02} are expected to be consistent with the AB system at better than 4 mmag.  \cite{Landolt07} showed that the Landolt magnitudes of various standard stars and the AB magnitudes are consistent to 6 mmag.  

Data from the Calan/Tololo survey \citep{CT} are not included in this analysis as the published number of comparison stars is quite small and a large fraction of the stars are below the $-30^{\circ}$ declination limit of the PS1 $3\pi$ survey.

\textbf{ CSP }- The basis for the CSP calibration is presented in \cite{Contreras10}.  The CSP optical follow-up campaigns were carried out with the Direct CCD Camera attached to the Henrietta Swope 1 m telescope located at the Las Campanas Observatory (LCO).  The survey observed in $ugriBV$ and the field-of-view of the observations is $ 8.7' \times 8.7'$.  Definitive measurements of the CSP filter throughput curves were carried out at the telescope using a monochromator and calibrated photodiodes (Rheault et al. 2010; \citealp{Stritzinger11}). 

CSP SN magnitudes are published in the native photometric system, defined by the Swope filter response functions of \citealp{Stritzinger11} and the primary standard BD+17. The CSP local standard magnitudes are published in the standard system, and we convert these magnitudes onto the CSP native system using the transformation equations provided.  As discussed in \cite{Stritzinger11}, we use the three different filter transmission functions for the CSP $V$ band (shown in Fig. 1) for the three periods of the CSP survey in which a different filter was used (labeled `CSP$_1$', `CSP$_2$', `CSP$_3$' for this analysis).

\textbf{ CfA4} - The basis for the CfA4 calibration is presented in \cite{Hicken12}.
The CfA4 data were obtained on the 1.2 m telescope at the Fred Lawrence Whipple Observatory (FLWO) using the single-chip,
four-amplifier CCD KeplerCam4. Observations were acquired in a field of view of approximately $11.5'\times11.5'$.  Cramer et al. (in prep) measured the FLWO 1.2 m KeplerCam $BVr^{\prime}i^{\prime}$ passbands using the monochromatic illumination technique initially described in \cite{Stubbs06}.  No atmospheric component is included in the Keplercam filter transmission curves.  As discussed in \cite{Hicken12}, the 1.2 m primary mirror deteriorated during the course of the CfA4 so that different transmission functions for various filters were recognized for different parts of the survey (labeled `CfA4\_1' and `CfA4\_2' in this analysis).  

\textbf{ CfA3} - The basis for the CfA3 calibration is presented in \cite{Hicken09a}.
The CfA3 sample was acquired on the F. L. Whipple Observatory
1.2 m telescope, mostly using two cameras, the 4Shooter camera and
Keplercam. The field of view of the observations was approximately $11.5'\times11.5'$.
\UBVRI filters were used on the 4Shooter (hereafter referred to as `CfAS')
while $UBVr^{\prime}i^{\prime}$ filters were used on Keplercam (hereafter referred to as `CfAK').  The 4Shooter \BVRI passbands are given in \cite{Jha06}.   The Keplercam  $UBVr'i'$ are measured in Cramer et al.\ (in prep.), as discussed in the CfA4 subsection above.

\textbf{ CfA2} - 
The basis for the CfA2 calibration is presented in \cite{Jha06}. The CfA2 sample was acquired on the F. L. Whipple Observatory
1.2 m telescope, with either the AndyCam CCD camera or the 4Shooter camera. Like for CfA3, the field of view was $11.5'\times11.5'$.  \UBVRI filters were used on both cameras.

\textbf{ CfA1} - 
The basis for the CfA1 calibration is presented in \cite{Riess99}. Like, CfA2, the CfA1 sample was acquired on the F. L. Whipple Observatory
1.2 m telescope, which has a field of view was $11.5'\times11.5'$.  \BVRI filters were used for all observations.  To include this sample in our analysis, we created our own catalogs of the stellar photometry presented by matching the catalogs given in the paper to the finding charts in the paper.

\begin{figure}
\centering
\epsscale{1.15}  
\plotone{{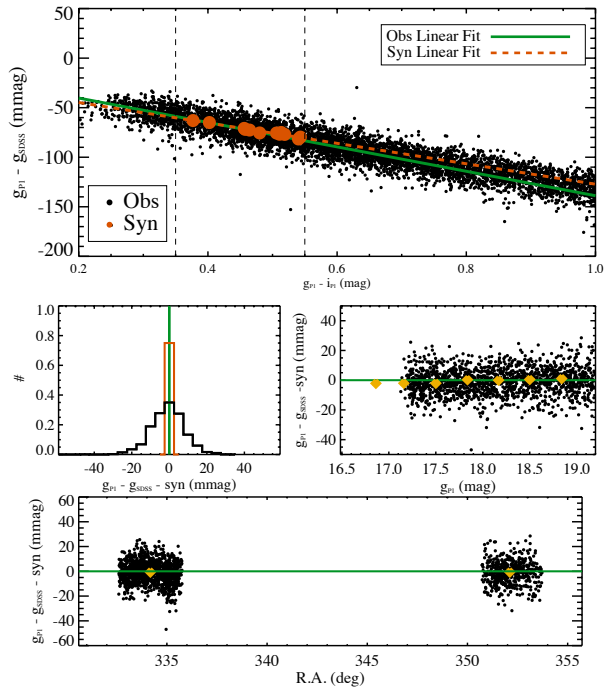}}
\caption{Visual representation of the steps of the Supercal method.  The top panel shows the synthetic and observed differences in brightness of stars observed with the $g_{\rm SDSS}$ and \gps band.  Observed stars are shown in black and synthetic stars are shown in red.  A linear transformation is fit to both the synthetic and observed trend. In the middle-left, a histogram is shown of the residuals from subtracting the synthetic trend from the observed stellar photometry.  The histograms are normalized to unity.  In the middle-right, the residuals as a function of magnitude are given. In the bottom panel, the dependency of the residuals on sky position is shown (bins given as yellow triangles).}
\label{fig:steps}
\end{figure}

\section{Cross Calibration}
\nobreak
\label{sec:cross}

Here we present a new calibration path: to directly tie all photometry catalogs to the homogeneous PS1-Ubercal catalog.  The PS1 catalog covers $3\pi$ of the sky with \about5 mmag relative calibration, and depth to \about22 mag.  In our new method, we compare observed color differences of stars in common to surveys with the synthetic transformations between the systems.  We then correct for these differences so that all samples are calibrated in a consistent manner.  The advantage of this new method is that there are much better statistics as we can use many of the stars in the catalogs that are directly used to determine zeropoints of the SN photometry. 

For a given star in which photometry is measured by two surveys in the AB system, the expected difference in magnitude can be expressed as:
\begin{equation}
O_{S^1_p}-O_{S^2_{p'}}=Y_{S^1_p}-Y_{S^2_{p'}}+\Delta_{S^1_p-S^2_{p'}}
\end{equation}
 where ${O_{S^1_p}}$~and~$O_{S^2_{p'}}$ are the observed magnitudes in passband $p$ and $p'$ in the two systems, $Y_{S^1_p}$, $Y_{S^2_{p'}}$ are the synthetic magnitudes (Eq. 2,3) of the systems and $\Delta_{S^1_p-S^2_{p'}}$ is the systematic discrepancy between how the calibration systems are defined.  Typically, $\Delta_{S^1_p-S^2_{p'}}$ is assumed to be 0; the main goal of this analysis is to test this assumption. $\Delta_{S^1_p-S^2_{p'}} \ne 0$ if there are errors in the measurements of the bandpass definitions, errors in the photometry of the star compared, errors in the synthetic spectrum of the star itself or errors in the AB offsets (Eq. 3) given for each system.  Given a number of standards with well-calibrated spectra, from Eq. 2 both $\Delta_{S^1_p-S^2_{p'}}$ and wavelength shifts to $p(\nu)$ and $p'(\nu)$ can be found by comparing the observed magnitudes of stars from two surveys.  

Our new approach consists of finding the differences of observed magnitudes of the same stars observed by two surveys, and comparing these differences with the expected differences from known passband definitions and a stellar library of synthetic spectra.  To do so, we fit a line to both the observed differences $\vec{O}_{S^1_p}-\vec{O}_{S^2_{p'}}$ of the two systems with color and the expected synthetic differences $\vec{Y}_{S^1_p}-\vec{Y}_{S^2_{p'}}$ of the two systems with color:
\begin{gather}
\vec{O}_{S^1_p}-\vec{O}_{S^2_{p'}}= \alpha_{O} (\vec{O}_{S^1_p}-\vec{O}_{S^1_{p2}}-c_0)+\beta_{O} \\
\vec{Y}_{S^1_p}-\vec{Y}_{S^2_{p'}}= \alpha_{Y} (\vec{Y}_{S^1_p}-\vec{Y}_{S^1_{p2}}-c_0)+\beta_{Y}
\end{gather}
where $\vec{O}$ and $\vec{Y}$ describe the vector of observed and synthetic magnitudes of all overlapping stars between two surveys (input values), $\alpha$ and $\beta$ are the linear component of the fit to the differences between the magnitudes for the observed and synthetic sequences (fitted values) and $c_0$ is a reference color that is chosen.  The linear fit is performed versus a transformation color ($\vec{O}_{S^1_p}-\vec{O}_{S^1_{p2}}$ or $\vec{Y}_{S^1_p}-\vec{Y}_{S^1_{p2}}$) as we expect differences in magnitude between surveys with similar passbands to depend on color.
If the passband shapes and edges are accurate but there is still a discrepancy in the zeropoints, then $\beta_{O}-\beta_{Y}=\Delta_{S^1_p-S^2_{p'}}$ at the reference color, $c_0$.  If the passband shapes and edges of either system are incorrect, then we should expect that $\alpha_{O} -\alpha_{Y} \neq 0$.  It is possible that $\alpha_{O} =\alpha_{Y}$ or $\beta_{O}=\beta_{Y}$ if systems are consistently wrong, though this scenario is unlikely.

To compare photometry from two different surveys, we follow a seven-step process in order to determine a single mean offset between the catalogs.
\begin{enumerate}
\item  Match the astrometric positions of stars observed by two surveys to $<1$ arcsec.  This ensures that stars will not be mismatched.  To avoid potential errors related to blending, only isolated stars, those with no other star with $m < 22$ mag within a 15 arcsec radius, are included in the sample.
\item  For a given band (e.g., $p$), or bands from different systems that are near each other in wavelength space, subtract the observed magnitudes from the two matched catalogs $(\vec{O}_{S^1_p}-\vec{O}_{S^2_{p'}})$.  Also determine the transformation color to use $(\vec{O}_{S^1_p}-\vec{O}_{S^1_{p2}})$ where $p$ and $p2$ are two passbands chosen to provide the strongest and most linear leverage on the color tilt of the magnitude differences.
\item  For the chosen bands, integrate the spectral library through the passbands of both systems to determine the synthetic magnitudes in those passbands.  For these library spectra, subtract the synthetic magnitudes to determine $(\vec{Y}_{S^1_p}-\vec{Y}_{S^2_{p'}})$.  Similarly, find the color $(\vec{Y}_{S^1_p}-\vec{Y}_{S^1_{p2}})$.
\item For the observed sequence of stars, adopt a magnitude cut to reduce the Malmquist bias from the PS1 faint stars included in the sample.  No magnitude cut is determined from the other samples as all catalog stars are used in the external analyses to determine the calibrate the SN photometry.  We also adopt a bright-end cut for PS1 magnitudes of [14.8,14.9.15.1,14.6] in \griz because of concerns about linearity brighter than these magnitudes \citep{Schlafly12}\footnote{Saturation is observed in PS1 magnitudes at [14.3,14.4.14.6,14.1] in \griz, and we conservatively add a half magnitude here.}.
\item For the observed sequence of stars, correct the stellar magnitudes for Milky Way reddening using the known positions of the stars and extinction values from \cite{Schlafly11} with a 2D sky map.  The extinction values are specific to each system.
\item Choose a specific color range (of e.g., $g-i$) of the catalog stars used in the analysis to reflect the same colors of the stars in the synthetic library.
\item Fit $\alpha_{O}$, $\beta_{O}$ and $\alpha_{Y}$, $\beta_{Y}$ to the observed and synthetic sequences respectively.  All photometric errors are propagated and we perform iterative $3-\sigma$ clipping to determine best fit values.

\end{enumerate}

An illustration of these steps is shown in Fig.~\ref{fig:steps}.  The offset $\Delta_{S^1-S^2_{pp'}}$ is determined from comparing the differences in $\beta$ at a reference color $c_0$ of the fitted lines to the synthetic and observed differences.  The statistical errors in either case includes both the error in the measurement of the linear fit to the observed and synthetic distributions.   We check the dependence of the offset on both brightness and R.A.\ but only make a cut on brightness to prevent a Malmquist bias.  We do not correct for a spatial bias any survey within this analysis, though this can be done in future analyses.

\section{Results}
\nobreak
\label{sec:consequences}
\subsection{Discrepancies Between Surveys}

We follow the procedure explained in the previous section to determine discrepancies between PS1 and each available catalog.  For each comparison, we fit the differences between observations from two different systems but with similar passbands, and attempt to remove the dependence on color.  Since the main purpose of this analysis is to more consistently tie all of the systems to the \textit{HST} Calspec system, we use the \textit{HST} Calspec library as our spectral library to determine the color transformation between the two systems.  All Calspec standards used in this analysis have a relatively small color range of $0.35 < g-i < 0.55$~mag.  Since we do not extrapolate beyond this color range, the number of stars used in the comparisons is limited.  This is further discussed in \S\ref{sys}.

For the higher-$z$ surveys, we choose the transformation color for these comparisons to be $\gps-\ips$ as we find this choice results in the smallest scatter in residuals.  We find other choices for the transformation color produce similar results, but with larger uncertainties.  Additionally, using $\gps-\ips$ rather than a color from another survey's observations allows us to be consistent in our comparisons (see Figure 2).  However, because of issues at blue wavelengths including the Balmer jump, in order to compare \gps to the $B$ band, we use the transformation color $B-i$.  In this case, we must perform a two-dimensional minimization as we find discrepancies in $B-\gps$ versus $B-\ips$, where the $B$ measurement comes from the comparison sample (see discussion of systematic uncertainties in \S\ref{sys}).

\begin{table*}[t]
\caption{Calibration discrepancies with PS1}
\begin{tabular}{llll|c|cccccc}
\hline \hline
Survey & Filt1 & Filt2 & NStar & \textit{HST} Off. ($\beta$)&  NGSL Off.*  ($\beta$) & $\textrm{Slope}_{Obs}$ ($\alpha$) & $\textrm{Slope}_{Syn}~(\alpha) $ & Slope Diff. & $d\alpha/d\lambda$ & $\Delta \lambda$ \\ 

~ & [PS1 ] & [O] & ~ & (mmag) & (mmag) & (mmag/mag) & (mmag/mag) & (mmag/mag) & $(\frac{\rm mmag}{(\rm mag \times \rm nm)})$ & (nm) \\

 \hline
CfA1 & $g$& $B$& 31 & $ ~ $ & $-15.5 \pm 18.6$ & $-302.2 \pm 54.2$ & $-334.3 \pm 9.1$ & $32.1 \pm 54.9$ & $9.4$ & $3.4\pm 5.8$   \\ 
CfA2 & $g$& $B$& 146 & $-1.3 \pm 13.5$ & $8.4 \pm 13.4$ & $-290.4 \pm 9.7$ & $-334.3 \pm 9.1$ & $43.9 \pm 13.3$ & $9.4$ & $4.7\pm 1.4$   \\ 
CfAK & $g$& $B$& 731 & $27.2 \pm 5.3$ & $30.9 \pm 5.3$ & $-317.6 \pm 8.0$ & $-315.3 \pm 12.2$ & $-2.3 \pm 14.6$ & $8.7$ & $-0.3\pm 1.7$   \\ 
CfAS & $g$& $B$& 396 & $30.8 \pm 5.3$ & $34.5 \pm 5.3$ & $-293.3 \pm 9.0$ & $-314.3 \pm 12.2$ & $21.0 \pm 15.1$ & $9.4$ & $2.2\pm 1.6$   \\ 
CfA4$_1$ & $g$& $B$& 855 & $19.8 \pm 5.7$ & $20.1 \pm 5.7$ & $-349.7 \pm 10.0$ & $-307.4 \pm 23.5$ & $-42.3 \pm 25.5$ & $8.7$ & $-4.9\pm 2.9$   \\ 
CfA4$_2$ & $g$& $B$& 912 & $-6.4 \pm 6.4$ & $-5.2 \pm 6.4$ & $-315.1 \pm 9.3$ & $-247.9 \pm 14.1$ & $-67.2 \pm 16.9$ & $9.3$ & $-7.2\pm 1.8$   \\ 
CSP1 & $g$& $B$& 251 & $4.6 \pm 4.7$ & $1.4 \pm 4.7$ & $-321.7 \pm 12.5$ & $-299.2 \pm 20.7$ & $-22.5 \pm 24.2$ & $12.8$ & $-1.8\pm 1.9$   \\ 
CSP1 & $g$& $g$& 362 & $-12.5 \pm 3.9$ & $-11.9 \pm 3.8$ & $-91.8 \pm 10.8$ & $-64.5 \pm 3.2$ & $-27.4 \pm 11.2$ & $5.8$ & $-4.7\pm 1.9$   \\ 
SNLS0 & $g$& $g$& 200 & $-11.5 \pm 4.1$ & $-17.9 \pm 4.1$ & $8.2 \pm 5.3$ & $18.0 \pm 1.3$ & $-9.8 \pm 5.4$ & $8.0$ & $-1.2\pm 0.7$   \\ 
SNLS1 & $g$& $g$& 272 & $-11.4 \pm 3.7$ & $-16.8 \pm 3.7$ & $18.0 \pm 6.4$ & $18.0 \pm 1.3$ & $-0.1 \pm 6.5$ & $8.0$ & $-0.0\pm 0.8$   \\ 
SNLS1 & $g$& $g$& 264 & $-10.4 \pm 6.5$ & $-14.7 \pm 6.5$ & $14.4 \pm 6.4$ & $18.0 \pm 1.3$ & $-3.6 \pm 6.5$ & $8.0$ & $-0.4\pm 0.8$   \\ 
SNLS3 & $g$& $g$& 568 & $-11.5 \pm 6.4$ & $-19.5 \pm 6.4$ & $12.5 \pm 5.4$ & $18.0 \pm 1.3$ & $-5.5 \pm 5.5$ & $8.0$ & $-0.7\pm 0.7$   \\ 
SDSS0 & $g$& $g$& 3300 & $-0.8 \pm 4.4$ & $-0.6 \pm 4.4$ & $-115.8 \pm 1.1$ & $-121.3 \pm 4.6$ & $5.5 \pm 4.7$ & $8.7$ & $0.6\pm 0.5$   \\ 
SDSS1 & $g$& $g$& 1525 & $-0.5 \pm 4.4$ & $-1.4 \pm 4.4$ & $-117.5 \pm 1.5$ & $-121.3 \pm 4.6$ & $3.8 \pm 4.9$ & $8.7$ & $0.4\pm 0.6$   \\ 
CfA1 & $r$& $V$& 22 & $ ~ $ & $-24.4 \pm 10.3$ & $-325.9 \pm 46.1$ & $-314.8 \pm 3.8$ & $-11.1 \pm 46.3$ & $5.4$ & $-2.1\pm 8.6$   \\ 
CfA2 & $r$& $V$& 120 & $-6.6 \pm 6.6$ & $-11.4 \pm 6.4$ & $-364.9 \pm 14.8$ & $-314.8 \pm 3.8$ & $-50.1 \pm 15.3$ & $5.4$ & $-9.3\pm 2.8$   \\ 
CfAK & $r$& $V$& 634 & $-1.1 \pm 3.6$ & $-2.6 \pm 3.6$ & $-357.4 \pm 10.8$ & $-348.8 \pm 5.9$ & $-8.6 \pm 12.3$ & $5.1$ & $-1.7\pm 2.4$   \\ 
CfAS & $r$& $V$& 339 & $2.1 \pm 3.8$ & $1.8 \pm 3.8$ & $-322.3 \pm 16.5$ & $-359.7 \pm 6.3$ & $37.4 \pm 17.7$ & $5.4$ & $6.9\pm 3.3$   \\ 
CSP$_1$ & $r$& $V$& 309 & $-8.4 \pm 3.6$ & $-12.6 \pm 3.7$ & $-393.1 \pm 11.5$ & $-362.0 \pm 6.6$ & $-31.1 \pm 13.3$ & $5.3$ & $-5.9\pm 2.5$   \\ 
CSP$_2$ & $r$& $V$& 309 & $-2.2 \pm 3.7$ & $-3.4 \pm 3.8$ & $-376.3 \pm 11.4$ & $-376.8 \pm 6.6$ & $0.5 \pm 13.2$ & $5.2$ & $0.1\pm 2.6$   \\ 
CSP$_3$ & $r$& $V$& 309 & $-6.2 \pm 3.9$ & $-9.2 \pm 4.0$ & $-395.1 \pm 11.6$ & $-380.0 \pm 7.2$ & $-15.1 \pm 13.6$ & $5.2$ & $-2.9\pm 2.6$   \\ 
CfA4$_1$ & $r$& $V$& 1038 & $-5.4 \pm 3.6$ & $-4.5 \pm 3.6$ & $-355.8 \pm 9.6$ & $-348.8 \pm 5.9$ & $-7.0 \pm 11.2$ & $5.1$ & $-1.4\pm 2.2$   \\ 
CfA4$_2$ & $r$& $V$& 1038 & $-5.9 \pm 3.7$ & $-5.1 \pm 3.7$ & $-355.8 \pm 9.6$ & $-346.4 \pm 5.9$ & $-9.4 \pm 11.2$ & $5.1$ & $-1.9\pm 2.2$   \\ 
CfA1 & $r$& $r$& 19 & $~$ & $-4.1 \pm 13.0$ & $177.7 \pm 46.4$ & $76.4 \pm 1.9$ & $101.3 \pm 46.4$ & $2.7$ & $36.9\pm 16.9$   \\
CfA2 & $r$& $r$& 119 & $5.2 \pm 5.1$ & $3.6 \pm 4.8$ & $13.8 \pm 20.7$ & $76.4 \pm 1.9$ & $-62.6 \pm 20.8$ & $2.7$ & $-22.8\pm 7.6$   \\
CfAS & $r$& $r$& 340 & $13.5 \pm 3.9$ & $14.8 \pm 3.9$ & $93.3 \pm 16.0$ & $79.7 \pm 2.3$ & $13.6 \pm 16.2$ & $2.7$ & $5.0\pm 5.9$   \\ 
CfAK & $r$& $r$& 627 & $-12.8 \pm 3.5$ & $-12.4 \pm 3.5$ & $6.5 \pm 10.9$ & $8.0 \pm 0.4$ & $-1.5 \pm 10.9$ & $3.1$ & $-0.5\pm 3.5$   \\ 
CfA4$_1$ & $r$& $r$& 1035 & $-9.1 \pm 3.4$ & $-4.9 \pm 3.3$ & $25.1 \pm 9.4$ & $8.0 \pm 0.4$ & $17.2 \pm 9.4$ & $3.1$ & $5.6\pm 3.1$   \\ 
CfA4$_2$ & $r$& $r$& 1035 & $-12.5 \pm 3.4$ & $-9.0 \pm 3.4$ & $13.5 \pm 9.3$ & $7.9 \pm 0.4$ & $5.6 \pm 9.3$ & $3.1$ & $1.8\pm 3.0$   \\ 
CSP1 & $r$& $r$& 309 & $-7.8 \pm 3.5$ & $-6.4 \pm 3.6$ & $1.7 \pm 10.2$ & $9.6 \pm 0.2$ & $-7.9 \pm 10.2$ & $3.2$ & $-2.5\pm 3.2$   \\ 
SNLS0 & $r$& $r$& 200 & $-8.2 \pm 3.5$ & $-7.2 \pm 3.5$ & $23.3 \pm 3.2$ & $20.6 \pm 0.6$ & $2.6 \pm 3.3$ & $3.0$ & $0.9\pm 1.1$   \\ 
SNLS1 & $r$& $r$& 272 & $-5.3 \pm 5.2$ & $-4.7 \pm 5.2$ & $23.5 \pm 5.7$ & $20.6 \pm 0.6$ & $2.8 \pm 5.7$ & $3.0$ & $0.9\pm 1.9$   \\ 
SNLS1 & $r$& $r$& 264 & $4.0 \pm 5.3$ & $4.0 \pm 5.3$ & $25.2 \pm 6.1$ & $20.6 \pm 0.6$ & $4.6 \pm 6.2$ & $3.0$ & $1.5\pm 2.1$   \\ 
SNLS3 & $r$& $r$& 569 & $-3.9 \pm 3.4$ & $-4.3 \pm 3.4$ & $20.7 \pm 5.2$ & $20.6 \pm 0.6$ & $0.1 \pm 5.2$ & $3.0$ & $0.0\pm 1.8$   \\ 
SDSS0 & $r$& $r$& 3300 & $-11.0 \pm 3.2$ & $-12.5 \pm 3.2$ & $-4.8 \pm 0.9$ & $-2.9 \pm 0.7$ & $-2.0 \pm 1.1$ & $3.3$ & $-0.6\pm 0.3$   \\ 
SDSS1 & $r$& $r$& 1525 & $-9.5 \pm 3.2$ & $-10.3 \pm 3.2$ & $-5.4 \pm 1.3$ & $-2.9 \pm 0.7$ & $-2.5 \pm 1.5$ & $3.3$ & $-0.8\pm 0.5$   \\ 
CfA1 & $i$& $I$& 19 & $~$ & $-4.7 \pm 22.3$ & $138.2 \pm 67.0$ & $87.6 \pm 3.8$ & $50.6 \pm 67.1$ & $1.9$ & $26.1\pm 34.7$   \\ 
CfA2 & $i$& $I$& 117 & $2.6 \pm 4.9$ & $3.0 \pm 4.8$ & $64.4 \pm 25.8$ & $87.6 \pm 3.8$ & $-23.2 \pm 26.1$ & $1.9$ & $-12.0\pm 13.5$   \\ 
CfAS & $i$& $I$& 319 & $9.2 \pm 4.4$ & $7.9 \pm 4.4$ & $88.4 \pm 18.6$ & $78.4 \pm 6.0$ & $9.9 \pm 19.6$ & $1.9$ & $5.1\pm 10.1$   \\ 
CfAK & $i$& $I$& 557 & $-1.5 \pm 3.6$ & $-0.8 \pm 3.6$ & $3.6 \pm 12.9$ & $12.2 \pm 0.7$ & $-8.5 \pm 12.9$ & $1.9$ & $-4.6\pm 6.9$   \\ 
CfA4$_1$ & $i$& $I$& 969 & $-0.1 \pm 3.6$ & $0.8 \pm 3.6$ & $17.4 \pm 10.3$ & $12.2 \pm 0.7$ & $5.2 \pm 10.3$ & $1.9$ & $2.8\pm 5.5$   \\ 
CfA4$_2$ & $i$& $I$& 969 & $0.4 \pm 3.6$ & $1.4 \pm 3.6$ & $17.4 \pm 10.3$ & $12.2 \pm 0.7$ & $5.2 \pm 10.3$ & $1.9$ & $2.8\pm 5.5$   \\ 
CSP$_1$ & $i$& $I$& 306 & $11.7 \pm 5.4$ & $11.3 \pm 5.4$ & $11.4 \pm 10.8$ & $14.4 \pm 0.9$ & $-2.9 \pm 10.8$ & $1.8$ & $-1.6\pm 6.0$   \\ 
SNLS0 & $i$& $i$& 200 & $-8.1 \pm 5.2$ & $-6.6 \pm 5.3$ & $22.6 \pm 5.1$ & $16.2 \pm 1.0$ & $6.3 \pm 5.2$ & $1.9$ & $3.3\pm 2.7$   \\ 
SNLS1 & $i$& $i$& 272 & $1.4 \pm 5.3$ & $1.8 \pm 5.2$ & $23.8 \pm 6.7$ & $16.2 \pm 1.0$ & $7.6 \pm 6.8$ & $1.9$ & $4.0\pm 3.6$   \\ 
SNLS1 & $i$& $i$& 264 & $3.4 \pm 5.3$ & $3.2 \pm 5.3$ & $31.4 \pm 5.7$ & $16.2 \pm 1.0$ & $15.2 \pm 5.8$ & $1.9$ & $7.9\pm 3.0$   \\ 
SNLS3 & $i$& $i$& 569 & $-0.8 \pm 3.4$ & $-1.7 \pm 3.4$ & $24.6 \pm 4.4$ & $16.2 \pm 1.0$ & $8.4 \pm 4.5$ & $1.9$ & $4.4\pm 2.4$   \\ 
SDSS0 & $i$& $i$& 3300 & $-5.1 \pm 3.4$ & $-5.6 \pm 3.4$ & $-2.9 \pm 1.0$ & $-7.2 \pm 0.4$ & $4.3 \pm 1.1$ & $1.8$ & $2.4\pm 0.6$   \\ 
SDSS1 & $i$& $i$& 1525 & $-8.6 \pm 3.4$ & $-9.7 \pm 3.4$ & $-4.6 \pm 1.4$ & $-7.2 \pm 0.4$ & $2.7 \pm 1.4$ & $1.8$ & $1.5\pm 0.8$   \\ 
CfA1 & $z$& $I$& 17 & $~$ & $2.4 \pm 20.7$ & $-22.3 \pm 54.8$ & $-105.1 \pm 6.0$ & $82.8 \pm 55.1$ & $1.9$ & $42.7\pm 28.4$   \\ 
CfA2 & $z$& $I$& 116 & $7.7 \pm 6.6$ & $8.2 \pm 6.6$ & $-127.2 \pm 21.3$ & $-105.1 \pm 6.0$ & $-22.1 \pm 22.1$ & $1.9$ & $-11.4\pm 11.4$   \\ 
CfAS & $z$& $I$& 363 & $2.1 \pm 5.5$ & $8.9 \pm 5.5$ & $-80.7 \pm 17.3$ & $-101.7 \pm 10.3$ & $21.1 \pm 20.1$ & $1.9$ & $10.9\pm 10.4$   \\ 
CfAK & $z$& $I$& 675 & $-5.6 \pm 7.0$ & $0.1 \pm 7.0$ & $-189.7 \pm 10.5$ & $-168.0 \pm 15.5$ & $-21.7 \pm 18.7$ & $1.9$ & $-11.7\pm 10.0$   \\ 
CfA4$_1$ & $z$& $I$& 1117 & $-8.4 \pm 6.8$ & $-0.5 \pm 6.7$ & $-168.7 \pm 9.4$ & $-168.0 \pm 15.5$ & $-0.7 \pm 18.1$ & $1.9$ & $-0.4\pm 9.7$   \\ 
CfA4$_2$ & $z$& $I$& 1117 & $-7.9 \pm 6.8$ & $0.0 \pm 6.7$ & $-168.7 \pm 9.4$ & $-168.0 \pm 15.5$ & $-0.7 \pm 18.1$ & $1.9$ & $-0.4\pm 9.7$   \\ 
CSP$_1$ & $z$& $I$& 337 & $4.9 \pm 7.2$ & $9.2 \pm 7.2$ & $-175.3 \pm 11.2$ & $-165.8 \pm 15.4$ & $-9.6 \pm 19.0$ & $1.8$ & $-5.3\pm 10.6$   \\ 
SNLS0 & $z$& $z$& 158 & $7.7 \pm 4.1$ & $11.6 \pm 4.1$ & $30.2 \pm 4.9$ & $17.4 \pm 2.0$ & $12.8 \pm 5.3$ & $1.6$ & $8.0\pm 3.3$   \\ 
SNLS1 & $z$& $z$& 210 & $5.6 \pm 5.7$ & $7.3 \pm 5.7$ & $30.4 \pm 8.2$ & $17.4 \pm 2.0$ & $13.0 \pm 8.5$ & $1.6$ & $8.1\pm 5.3$   \\ 
SNLS1 & $z$& $z$& 264 & $10.8 \pm 5.7$ & $11.4 \pm 5.7$ & $40.0 \pm 5.8$ & $17.4 \pm 2.0$ & $22.6 \pm 6.1$ & $1.6$ & $14.1\pm 3.8$   \\ 
SNLS3 & $z$& $z$& 468 & $4.3 \pm 4.2$ & $6.2 \pm 4.2$ & $26.6 \pm 5.4$ & $17.4 \pm 2.0$ & $9.1 \pm 5.7$ & $1.6$ & $5.7\pm 3.6$   \\ 
SDSS0 & $z$& $z$& 3295 & $16.4 \pm 3.9$ & $16.6 \pm 3.9$ & $37.3 \pm 1.6$ & $27.5 \pm 2.9$ & $9.8 \pm 3.3$ & $1.6$ & $6.3\pm 2.2$   \\ 
SDSS1 & $z$& $z$& 1525 & $16.5 \pm 3.9$ & $16.8 \pm 4.0$ & $35.5 \pm 2.1$ & $27.5 \pm 2.9$ & $8.0 \pm 3.6$ & $1.6$ & $5.2\pm 2.3$   \\ 

\hline
\end{tabular}
Offset and slope differences between each system and PS1.  The second and third column show the PS1 filter and comparison system filter.  The fourth columns shows the total number of overlapping stars between $0.3<g-i<1.0$.  The fourth column shows the offset found when using the \textit{HST} Calspec library to find a nominal calibration offset between the systems.  These offsets should be added each survey's magnitudes to agree with PS1.  Columns 5-10 show both the calibration offset and difference in predicted slopes when using the NGSL library for the spectral transformation.  Columns 9 and 10 show how to convert the difference in predicted and recovered slopes of the transformation to a change in the mean wavelength of the comparison filter.  The offset given when using \textit{HST} Calspec library is for a color range of $0.35<g-i<0.55$ while the offset and slope when using using the NGSL library is for the color range of $0.3<g-i<1.0$.   CfA1 values are missing due to lack of comparison stars for a small color range. \end{table*}

 We present the results from the offsets ($\Delta_{S^1_p-S^2_{p'}}$) found after removing the color transformation (see Fig. 2) in Fig. 3.  Here, $\Delta_{S^1_p-S^2_{p'}}=\beta_O-\beta_Y$ and we analyze the very small color range of \textit{HST} Calspec standards such that $\alpha_O-\alpha_Y$ is insignificant.  Offsets with respect to $\griz$ are shown for each filter of each system.  The errors shown include systematic uncertainties which are discussed in the next section.  We separate SNLS into its 4 deep fields and SDSS-II into the two parts of Stripe82 that overlap with PS1 Medium Deep fields and one part for the entire Stripe82.  The largest deviations are seen in the comparison between $g$ and $B$ bands.  Scatter of these offsets is around $2~\textendash~3\%$.  The offsets relative to the $\rps$ and $\ips$ bands are generally $<10$ mmag though there appears to be a systematic offset between PS1 and other surveys in the $r$ band.  The offsets found relative to $z$ band are slightly larger; in particular, there is a \about15 mmag offset between PS1 and SDSS.  The errors shown in Fig.~3 account for both the uncertainty from the observed magnitudes of stars as well as the uncertainty in the transformation from the synthetic magnitudes of the library standards.  The uncertainties are largest for the comparison between \gps and the low-$z$ systems' $B$ because there is the most scatter in the transformation between these two filters and the number of stars used for these comparisons is typically not high.  

For SNLS, we are able to perform these comparisons for each of their deep fields independently and find scatter $<5$ mmag which shows both the precision of the PS1 Ubercal and of the SNLS calibration.  Similarly, we can verify the relative calibration of the PS1 Ubercal in the comparisons with SDSS for the two Medium Deep fields that overlap with Stripe82.  The largest difference between the two Medium Deep fields when comparing against SDSS is $4$ mmag.  Overall the differences seen here between the SDSS and SNLS calibration is \about10 mmag in each band, depending on the field.  This is roughly within the errors given for the joint analysis between the two surveys \citep{Betoule12}.  

In the Table 4, we present the offsets of $\beta_O-\beta_Y$ using the \textit{HST} Calspec library at $c_0=(g-i)_0=0.45$ mag, and both $\beta_O-\beta_Y$ and $\alpha_O-\alpha_Y$ when using the NGSL library and a large enough color range to measure the slopes of the color transformations.  We also give the number of stars common to PS1 and each survey for each comparison.  As discussed below, we cannot accurately quantify the systematic uncertainties of the measurements of the slope difference  $\alpha_O-\alpha_Y$, so the information given here is solely for future study.  However, we note the large magnitude of the differences in slopes for some of these comparisons.  For better understanding, we convert the difference in slope to a nominal difference in the mean effective wavelength for the filter used in the given survey.  Both the dependence of the slope on mean effective wavelength as well as the corresponding shift necessary to bring the calibration into agreement with PS1 is given.  Many of these values are larger than the quoted systematic uncertainties in Table 1.

The largest discrepancies seen in the $B$ band of the low-$z$ systems are most likely due to the use of  BD+17 as the primary calibration standard.  \cite{Bohlin15} show that the luminosity of this standard has varied over the last two decades by \about4$\%$ and using it to anchor the calibration will result in additional systematic biases.  If we instead recalibrate the low-z surveys based on the Calspec standards P177D or P330E, as suggested by \cite{Bohlin15}, we find that the change in B magnitudes is negligible, but the change in $VRI$ magnitudes is $-30$ mmag.  The net result would be a consistent $30$ mmag offset in every filter when comparing the low-$z$ \BVRI to the higher-$z$ systems.  Therefore it appears likely that there is indeed an error in the low-$z$ systems of $\Delta (B-V)=30$ mmag and possibly an additional gray offset of a similar magnitude.

\begin{figure*}
\centering
\epsscale{1.15}  
\plotone{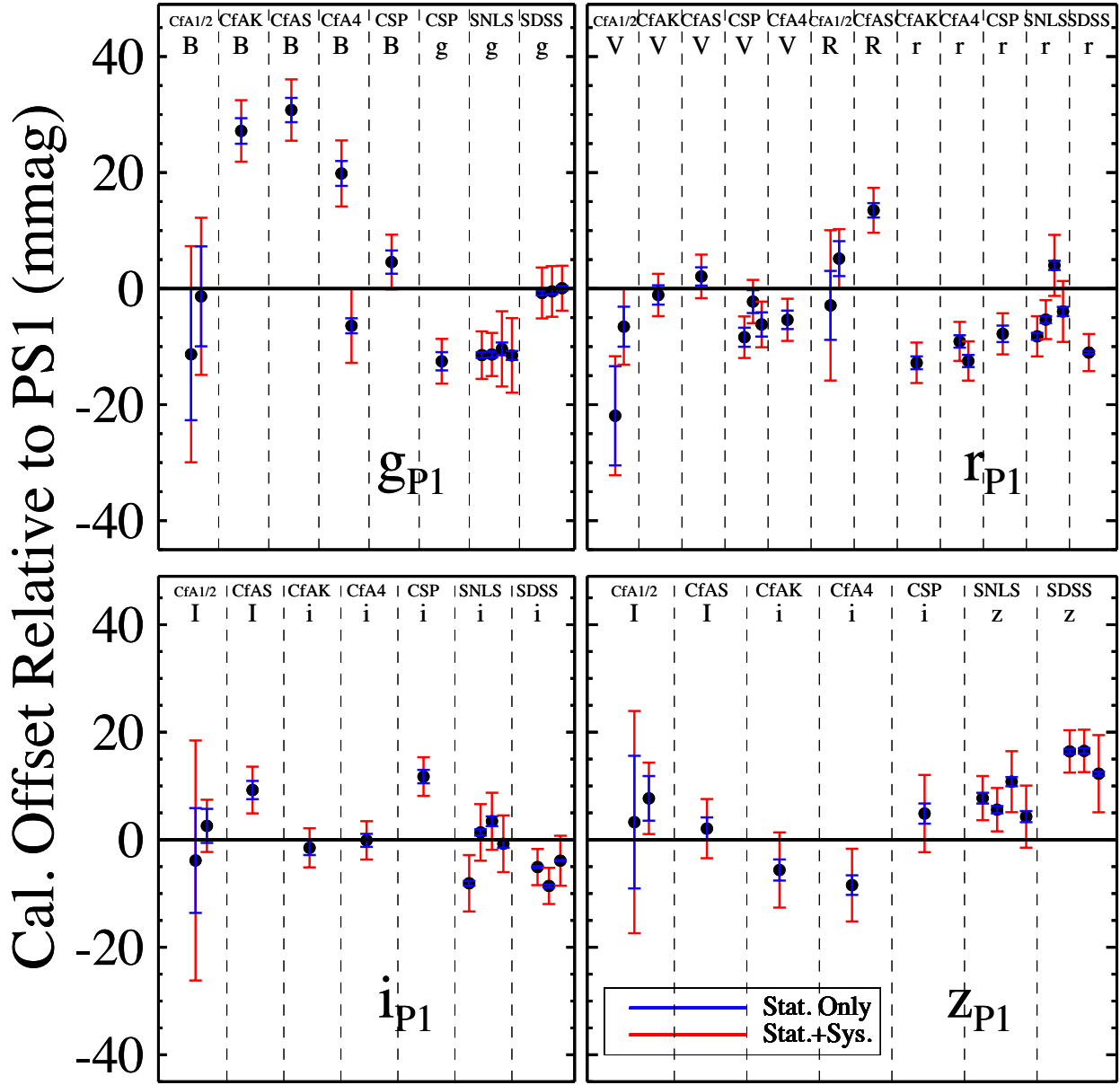}
\caption{The deviations from agreement with PS1 calibration for all surveys and filters.  The surveys shown are CfA1, CfA2, CfAS, CfAK, CSP, CfA4, SNLS, SDSS.  The system and filter are listed for each comparison..  The uncertainties shown here include both the statistical and systematic uncertainties from the Supercal process.  The offsets shown here should be added to each system's magnitudes to agree with PS1.}
\label{fig:Sys}
\end{figure*}
One can adjust the PS1 zeropoints such that the PS1 calibration is aligned to any other system.  Doing this, we can then measure the discrepancies between the recalibrated PS1 system and all other systems.  For example, correcting the PS1 system to align with SNLS requires applying offsets of $-12, -6, +1, +7$ mmag from \griz, respectively.  We can then determine the discrepancies between all other systems and the PS1+SNLS system.

We release online\footnote{www.kicp.uchicago.edu/$\sim$dscolnic/supercal/} all of the data used to make each comparison, and all iterations of versions of Figure 2 for all of the comparisons done in this analysis.  We also include online all of the transmission functions for each of the systems used in this analysis, as well as all of the zeropoints for each filter.

\subsection{Systematic Uncertainties}
\label{sys}
The largest systematic uncertainties in this approach are: the consistency of the PS1 Ubercal, accuracy of the spectral library, reddening of the stars from dust and PS1 linearity.  A smaller systematic uncertainty is the variation of the filter functions with radial position on the focal plane, though for both PS1 and SNLS, knowledge of this variation is used to correct the magnitudes based on their radial position.  For all other systems, this variation is expected to be negligible.  A summary of the dominant systematic uncertainties is given in Table 5.

The systematic uncertainties from the PS1 Ubercal are given in Finkbeiner et al. (2015, submitted).  By comparing the relative calibration of SDSS and PS1, Finkbeiner et al. 2015 finds that for a given sky position, there are systematic uncertainties less than {9,7,7,8} mmag in $griz$.  These uncertainties include both the systematics from SDSS and PS1, not only PS1.  A fair upper limit for the systematic uncertainty due to PS1 is roughly half these uncertainties: {5,4,4,4} mmag.  These uncertainties are further reduced for the PS1 Medium Deep fields due to the large number of observations of these fields and are expected to be around 3 mmag in each filter.  Since most of the comparisons between PS1 and other surveys involve stars from multiple fields, we estimate the systematic uncertainty in the PS1 Ubercal photometry to be the quadrature sum of an error floor of 3 mmag in addition to the filter-specific uncertainty ({5,4,4,4} mmag) divided by the square root of the number of fields.  

The systematic uncertainty associated with the spectral library used for the synthetic transformations can be determined by comparing the transformations using multiple independent spectral libraries.  We analyze stellar spectra from 5 different libraries: the \textit{HST} Calspec library (version 005)\footnote{Differences in absolute flux of Calspec standards between version 003 and version 005 are on the 8 mmag level}, the NGSL spectral library \citep{Heap07}, the INGS spectral library (Pickles et al. in prep), the `Pickles Atlas' \citep{Pickles98} and the Gunn-Stryker library \citep{Gunn83}.  We must make specific cuts for each library to properly alleviate systematic biases and optimize the most consistent comparisons.  For the \textit{HST} Calspec library, the NGSL library and the INGS library, we explicitly only include solar analog stars ($-0.6<\rm{[O/H]}<0.3$).  This cut allows for the most direct comparison between the majority of main sequence stars observed and the stars from the synthetic libraries.  For the \textit{HST} Calspec library, we only include standards that have been recently calibrated with WFC3 so that only the latest updates are used.  Because of known flux biases with distance from slit center in the NGSL observations\footnote{More information found here: https://archive.stsci.edu/prepds/stisngsl/}, we only include standards that were observed within 0.5 pixels of the slit center.  There is significant overlap between the INGS and NGSL libraries as the INGS library uses many NGSL spectra and includes its own correction for slit-loss in the NGSL observations.  We do not exclude any spectra from the Pickles library or the Gunn-Stryker library.

\begin{figure*}
\centering
\epsscale{\xScale}  
\plottwo{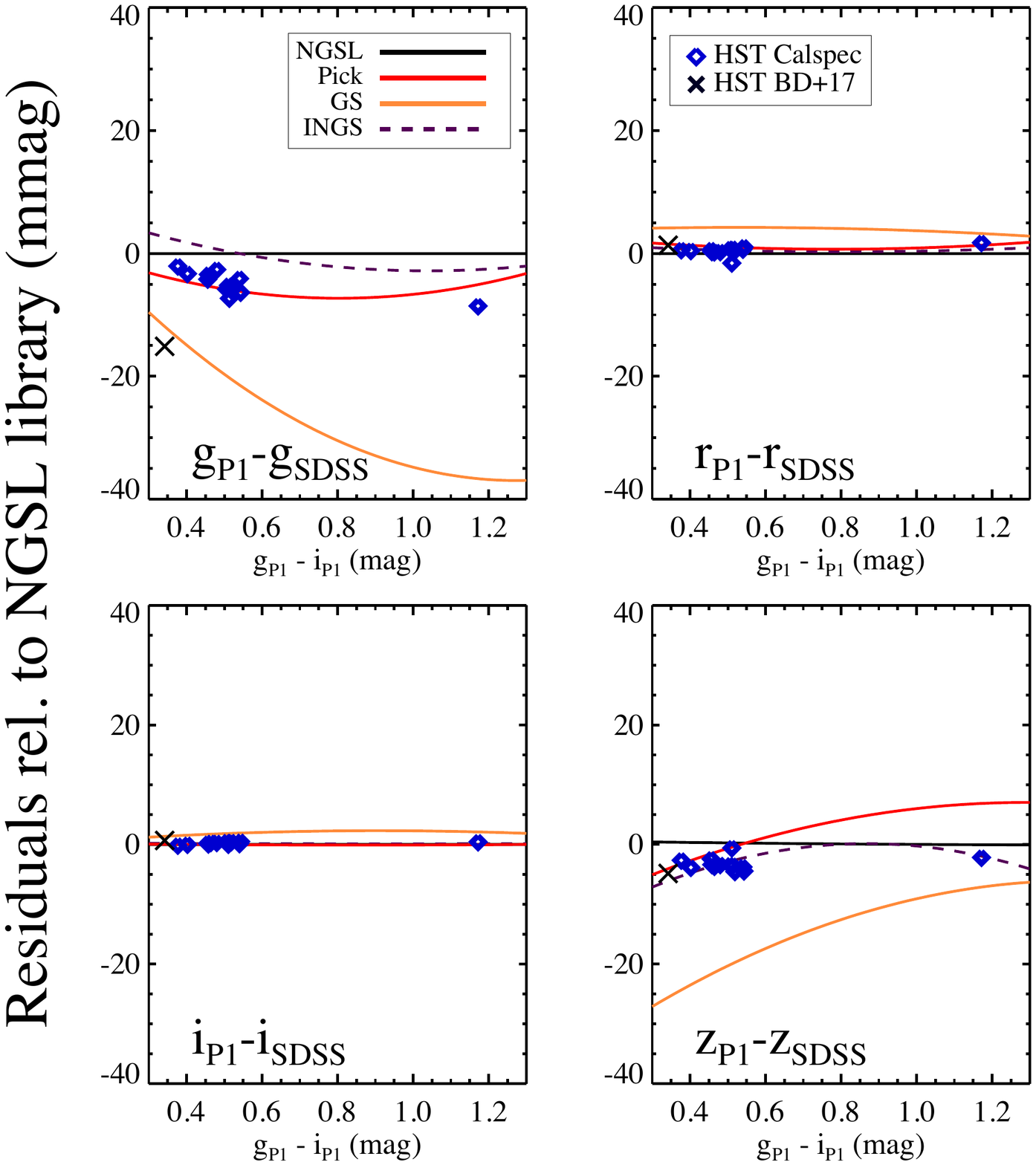}{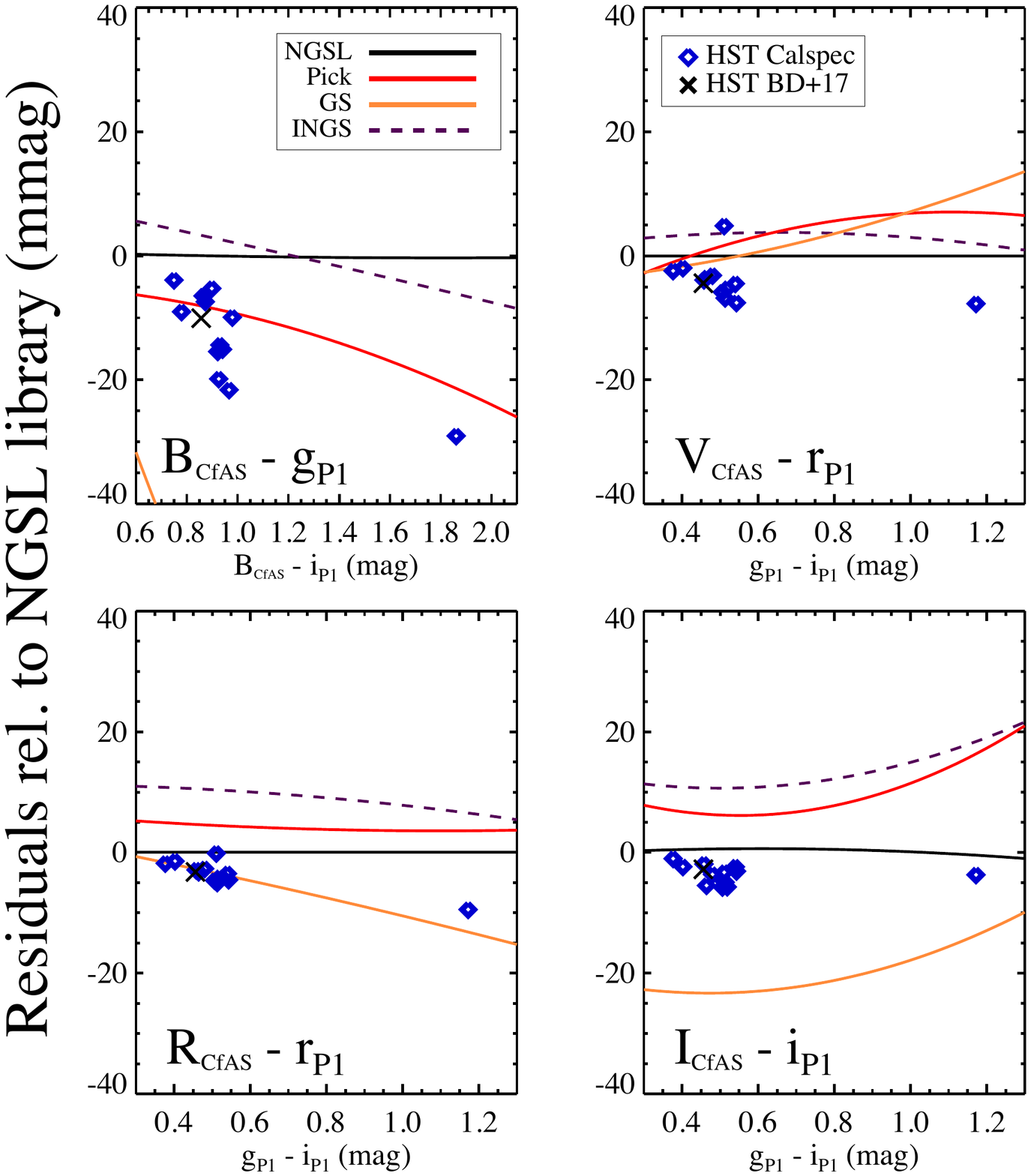}
\caption{Left: Residuals of the synthetic transformations between PS1 and SDSS when using different spectral libraries.  The $\gps-\ips$ color is used as the transformation color for these systems.  The transformation with the NGSL library is used for the baseline trend and is removed for each case.  The \textit{HST} Calspec standards are represented by points on the plots.  Right:  Residuals of the synthetic transformations between PS1 and CfAS when using different spectral libraries.  The colors $B_{\rm CfAS}-\ips$ and $\gps-\ips$ are used as the transformation colors for these systems. }
\label{fig:libraries}
\end{figure*}

Examples of differences between these libraries are shown in Fig.~\ref{fig:libraries} where we find the synthetic transformations between PS1 and SDSS as well as PS1 and CfAS.  We only show the transformations for a color range of $0.3<\gps-\ips<1.5$ as that is the only part of the main sequence where many stars can be found and there is the smallest amount of scatter in the transformation.  We see that there is clearly less of a dependence on the spectral library for the transformation between PS1 and SDSS as the mean effective wavelengths and wavelength range of these filters are much closer.  However, the dependence can be quite significant when comparing filters between PS1 and CfAS; separation between the mean effective wavelength of filters from these systems can be $>200~\rm \AA$.  

Since we chose to use the HST Calspec library as our primary spectral library, we determine the systematic uncertainty in the synthetic transformation between systems by instead using the \textit{HST} NGSL library.  As the NGSL library was also acquired using \textit{HST}, it is a useful comparison to the \textit{HST} Calspec library.  Included in the uncertainties given in Fig. 3, we find discrepancies in the offsets when using these two libraries to be in the range up to 4 mmag.  Differences between the \textit{HST} Calspec library and other libraries, like the Gunn-Stryker library or the Pickles library, can be significantly larger.  We are limited to use a small color range when comparing systems due to the small amount of standards in the \textit{HST} Calspec library.  We can increase this color range by using the NGSL library to extend from $0.35<\gps-\ips<1.0$.  However, since we do not have multiple \textit{HST} Calspec standards in this larger color range, we cannot use two libraries to assess what an appropriate systematic uncertainty would be for this range.

The impact of dust depends on whether there is a difference in the dependence on color between the dust reddening vector and temperature vector.  This issue can be largely removed by accounting for the extinction in the field that each star is located in.  The uncertainty in the extinction values is discussed in S14.  From \cite{Schlafly11}, we can safely assume that since the stars used in this analysis are removed from the galactic plane, all of the dust represented by the MW extinction value is between us and the stars used for this analysis, and any uncertainty in this assumption is included in the total systematic uncertainty of the extinction values.  We propagate the uncertainty of extinction values through the Supercal process and find that for an extinction of $E(B - V)=0.1$ mag, we can expect systematic uncertainties of $2,0.5,0.5,1.0$ mmag in $griz$ respectively.   

\begin{figure}
\centering
\epsscale{1.15}  
\plotone{{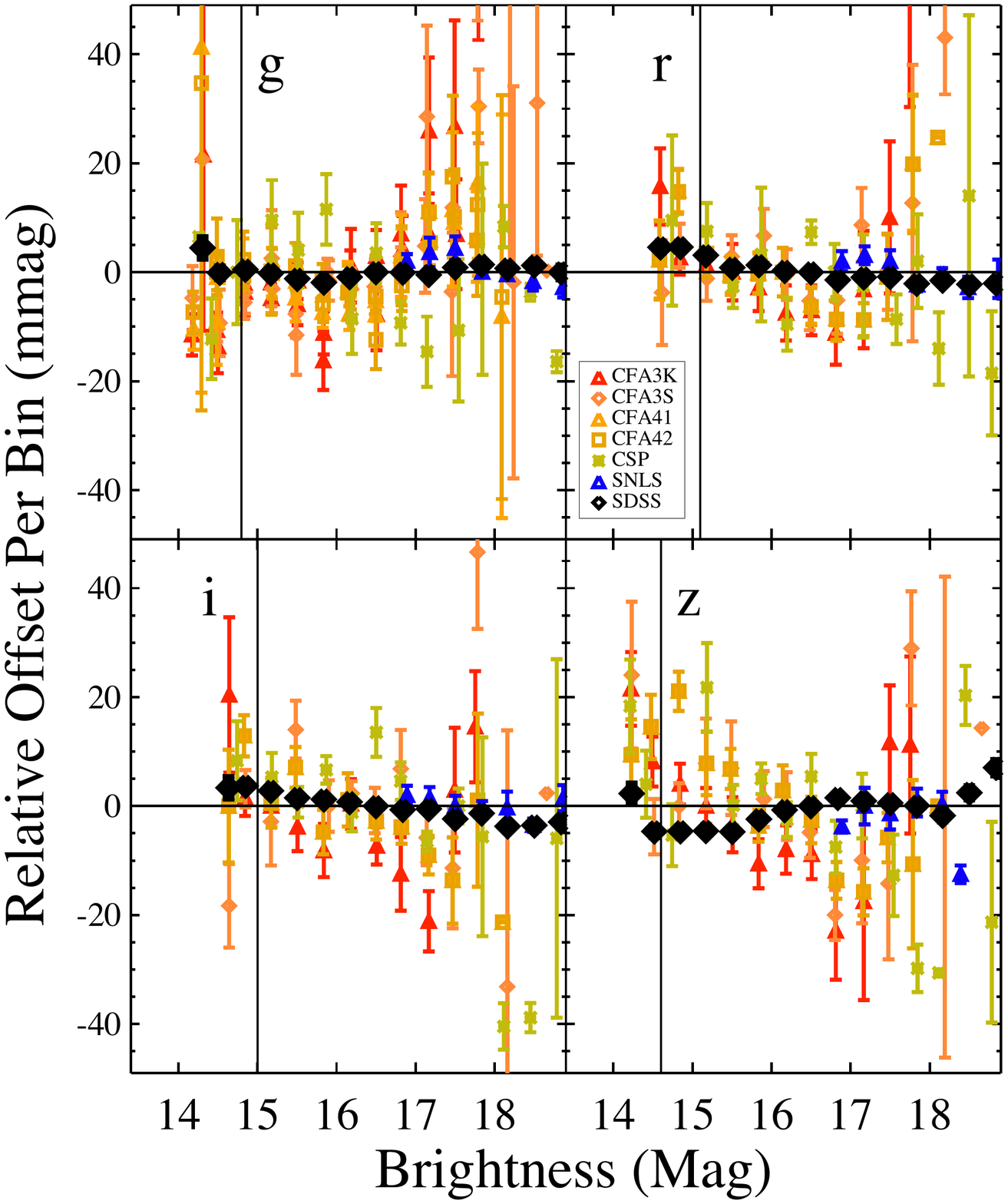}}
\caption{Relative offsets between stars in the PS1 sample and other samples as a function of apparent brightness in \griz.  The offsets shown are relative to the nominal Supercal correction given in Fig. 3.  Comparisons for each survey are shown, and in black the offsets for SDSS are highlighted to quantify potential systematic uncertainties from the PS1 non-linearity.  Vertical lines show where the PS1 sample is cut on the bright end.}
\label{fig:bias}
\end{figure}

\begin{figure}
\centering
\epsscale{1.15}  
\plotone{{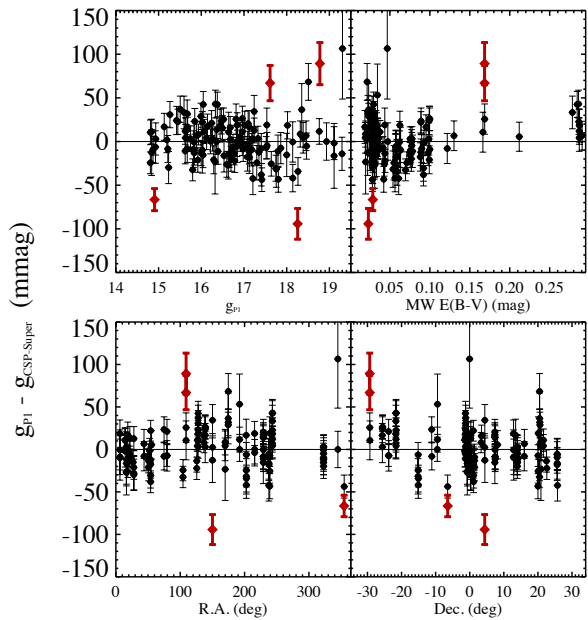}}
\caption{Relative offsets between stars in the CSP sample and PS1 as a function of apparent brightness in \gps, Milky Way reddening, R.A. and Dec.  The offsets shown here reflect the values after the Supercal correction from Fig. 3 is applied for the $g_{\rm CSP}$ filter.  Points in red are excluded from the offset calculation as part of the iterative 3-$\sigma$ clipping procedure. }
\label{fig:bias}
\end{figure}

\begin{table}[ht]
\caption{Main Systematic Uncertainties in the Supercal Process}
\begin{center}
\begin{tabular}{l|c}
\hline \hline
Systematic: & Uncertainty \\
\hline
Spectral Library & $0<x<4$ mmag \\
Dust & $0.5<x<2$ mmag \\
PS1 Ubercal & $3<x<6$ mmag \\
PS1 Non-linearity & $0<x<3$ mmag \\
\end{tabular}\\
Top systematic uncertainties in the Supercal process.  A range of uncertainties is given here, while the individual systematic uncertainty for each system/bandpass is included in the discrepancies shown in Figure 3 and given in Table 4.\\
\end{center}
\end{table}

We investigate in Fig. 5 the systematic uncertainties due to non-linearity of the PS1 stellar magnitudes.  Here we remove the nominal $\beta$ offset shown in Fig. 3 for each survey to probe possible biases with magnitude.  Overall, we choose to quantify the possible non-linearity due to PS1 by finding the trend for each filter when comparing with SDSS.  This is roughly 0,2,3,5 mmag for $\griz$ respectively over a 3 mag range from \about15-18 mag of each filter.  It is unclear how much of this non-linearity to attribute to PS1, since comparisons with SNLS and other surveys do not appear to favor the same biases with magnitude that SDSS does.  Therefore we conservatively claim that half of this non-linearity is due to SDSS and half to PS1.  From this plot, we also see significant trends at the faint magnitudes for the low-$z$ surveys.  Likely, these trends are indicative of either selection biases or poor photometry of fainter stars in the low-$z$ surveys.  These biases may also affect the SN photometry, though this will require analysis in a future study.

We limit the systematic uncertainties to ones that are based on PS1 and the methodology used here, and not of any of the other individual surveys.  Exploring the comparisons between PS1 and other surveys, many interesting pathologies can be found.  For example, there are certain fields of stars in the low-$z$ surveys that appear to have a significantly different offset when comparing to PS1 than the majority of fields of these surveys.  One example can be seen in the comparison between $\gps$ and $g_{\rm CSP}$ in Fig. 6.  While we find a negligible dependence of the relative offsets based on magnitude, we do see some dependence on MW dust reddening.  However, this trend depends strongly on a handful of fields with high reddening, and it is unclear if the trend is caused by the reddening or if those particular fields happen to be discrepant for other reasons.  Overall we see on the order of 5~\textendash~10 mmag trends with brightness, dust or R.A./Dec. for CSP as well as many of the other low-$z$ surveys.  The 5~\textendash~10 mmag trends represents a rough estimate of the systematic uncertainties in these samples.

\subsection{Comparing SN Photometry}

A more direct method to determine calibration discrepancies between various filters with the ultimate goal of consistent SN photometry would be to compare the photometry of SN observed by multiple surveys.  Unfortunately, this approach is limited by both the statistics and the methodology --- comparing photometry of SN is much more difficult as the SN\,Ia features are relatively deep and broad relative to stellar absorption, yet narrow compared to the width of a filter.  \cite{Mosher12} compared photometry of 9 SN\,Ia observed by both SDSS and CSP, though argued that 4 of these SN could not be used as their cadences were not satisfactory or the SN were not typical SN Ia, which limited the comparison.  Of the remaining $5$ SN, \cite{Mosher12} found that the photometry of each individual SN agreed to better than 1\%, though the sample scatter is up to $8\%$ and there is no single, consistent offset for all SN.  \cite{Betoule14} performs a similar comparison to that of \cite{Mosher12} when trying to compare CfA3 and CSP.  Their sample has 17 SN,  and they find offsets in $BVri$ of $5\pm4$,$-9\pm3$,$24\pm4$,$3\pm12$ mmag with scatter of $42,21,39,49$ mmag respectively.  No systematic uncertainties are given for this approach, and we find using Supercal that these differences are $10\pm8,3\pm9,4\pm9,-4\pm6$.  A larger sample is needed to understand if there are errors beyond those seen in calibration, such as those due to image subtraction or astrometric registration.  An alternative way to assess whether Supercal improves the consistency of these two surveys is to look at the agreement in the measured distances for the same SN using photometry from separate surveys.  We find without Supercal, the difference in distances between the CSP and CfA3 surveys for 23 SN is $0.031\pm0.026$ mag and after Supercal the difference is $-0.012\pm0.026$ mag.  More statistics will be needed for a more robust diagnostic.

\subsection{Effects on Recovered Cosmology}

We may choose to remove the calibration discrepancies found between all the systems so as to create a more uniformly calibrated sample.  We are able to force all systems to be relatively calibrated in a consistent manner with a given system or the average calibration from multiple systems.  For the average solution, we use the calibration from PS1, SNLS and SDSS, which are the most recently calibrated systems, to determine a joint solution of the baseline-Supercal calibration.  We use the offsets shown in Fig. 3, and weight them by the systematic uncertainties given in Table 1.  After doing so, we find average offsets from the PS1 calibration of $-5\pm3,-7\pm3,-2\pm2,6\pm4$ mmag where the uncertainties given are from both the systematic uncertainties given in Table 1 combined with the uncertainties from the Supercal process.

After making the systems consistent, we can then redetermine the cosmological parameters derived from the full sample.  The light curves of all SN are fit with the SNANA package \citep{SNANA} so that the fits are consistent with \cite{Betoule14}; the same SALT2 model and host-mass Hubble residual step of $0.06$ mag are applied.  Further information about light-curve fitting, distance bias corrections and host masses will be discussed in the next PS1 cosmology analysis (Scolnic et al. in prep).   The number of SN for each system, after quality cuts explained in \cite{Betoule14}, is: CfA1 (10), CfA2 (18), CfAS (33), CfAK(58), CfA4\_1 (34), CfA4\_2 (9), CSP(32), SDSS (359), PS1 (111), SNLS (235).  Differences in the recovered cosmology, when including only CMB constraints from \cite{Planck14}, are shown in Table 6 for the unaltered calibration, as well as when the calibration is forced to agree with the SDSS, SNLS, PS1 or the average calibration.  The weighting of each sample is determined by the error shown in Figure 3 for each system, and we also include calibration errors from the SALT2 model and from the HST AB system as part of the weighting covariance matrix.  We find that values for $w$ may change up to $\Delta w\sim-0.040$ depending on which primary calibration is used.  For the average Supercal solution based on SDSS, SNLS and PS1, the change in $w$ is $-0.026$.  The primary cause of these changes is due to the uncertainty in the B and V bands which affects the low-$z$ SN.  For the average Supercal calibration, we show the mean Hubble residual for each SN subsample in Fig. 7 (top) and the difference in distances as a function of redshift when the Supercal correction is applied relative to when it isn't in Fig. 7 (bottom).  The Supercal solution appears to significantly improve the agreement between the low-$z$ samples and the $\Lambda$CDM model.  We find mostly positive Hubble residuals from the $\Lambda$CDM model due to the host mass correction and the distance-bias correction (for latter correction, see Figure 5 in \cite{Betoule14}; more discussion in upcoming Scolnic et al. in prep). The net change in distances for each subsample is up to $0.055$ mag for the low-$z$ systems, but only $1\%$ for the higher-$z$ systems.

The Supercal process has a small effect on the distance scatter of the joint sample.  For the entire joint sample shown in Fig. 7, for each of the five cases shown in Table 6, we find a relative $\Delta \chi^2$ of $[0.0, -2.3, +6.7, -9.9,+0.3]$ respectively for 906 SN.  However, if we only measure the improvement in $\chi^2$ for the SNe with $z<0.1$, we see a more significant improvement in each case: $\Delta \chi^2$ of $[0.0, -3.2, -10.7,-7.2,-11.2]$ respectively for 225 SN.  The effect of Supercal is obviously much larger for the low-z sample as offsets between PS1, SDSS and SNLS are small and the calibration of SDSS and SNLS has already been connected in \cite{Betoule12}.  In a full cosmology analysis, SALT2 should be retrained with the optimal Supercal solution and simulated biases of distance residuals with color should be removed \citep{S14a}.  These improvements should better show the impact of the Supercal corrections.
\begin{table}[ht]
\caption{Differences between the various surveys}
\begin{tabular}{l | c c |}
\hline \hline
Primary Calibration & $\Delta w$ & $\Delta \Omega_M$ \\
\hline
No-Correction & $0.000 \pm0.052$ & $0.000\pm0.017$ \\
Supercal--Avg & $-0.026\pm0.051$ & $-0.005\pm0.015$\\
Supercal--PS1 & $-0.040\pm0.055$ & $-0.015\pm0.015$ \\
Supercal--SDSS & $+0.010\pm0.052$ & $+0.005\pm0.017$ \\
Supercal--SNLS & $-0.032\pm0.050$ & $-0.007\pm0.016$ \\

\hline
\end{tabular}

Differences in the recovered cosmology when calibration is forced to agree with that of a particular survey.  The no-correction value is given when no change to any calibration is made.  Errors given represent the total systematic error of the best fit cosmology and include the uncertainties from the Supercal process for each filter/system.\\
\end{table}

\begin{figure}
\centering
\epsscale{1.15}  
\plotone{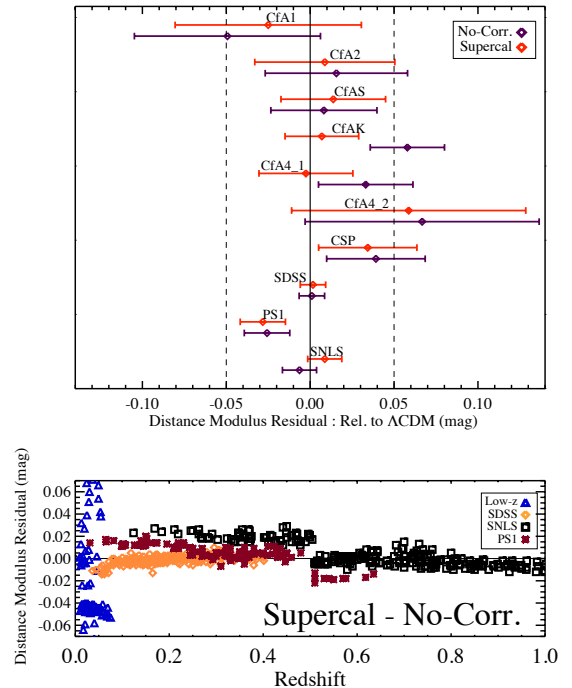}
\caption{(Top) Mean relative Hubble residual offsets to the $\Lambda$CDM model for SN observed by each system analyzed.  The points in black represent the residuals when no change to the the calibration systems are made, and in red, they represent the offsets after we apply the average Supercal correction.  (Bottom) The change in distances due to forcing all calibration to agree to the average Supercal solution.  The values are such that $\Delta \mu =\mu_{\rm Supercal}-\mu_{\rm No Corr.}.$  SN from different samples are represented by different symbols and colors as indicated in the legend.}
\label{fig:Sys}
\end{figure}

\section{Discussion}
\label{sec:discuss}

The accuracy of the Supercal technique presented in this paper depends on the number of stars common to PS1 and the comparison system, as well as the difference in effective wavelengths for the filters that are being compared.  While PS1 has covered nearly the limit of its observable sky, it would still be possible to increase the PS1 statistics of bright stars by observing at shorter exposure times.  This could help the absolute calibration of PS1 as well as the Supercal process.  It is no longer possible for most of the other surveys analyzed here to obtain more observations of stars for other surveys, though some could work to release more of the data already acquired.  A clear path to improved accuracy would be to redefine the absolute calibration of the low-$z$ surveys on multiple standards observed by \cite{Landolt92}, rather than BD+17.  In the current scheme, the accuracy of the calibration of the low-$z$ survey is reliant on the accuracy of the observations of BD+17 by both \cite{Landolt92} and \textit{HST}.  As BD+17 is a binary star, this strategy is sub-optimal.  

Furthermore, a complete understanding of the impact of these relative differences between surveys can be understood only after the light-curve models are retrained with the new calibration.  As the light-curve training is primarily based on low-$z$ SN, any systematics in the calibration of these SN has an appreciable effect on the fitted SN distances.

The Supercal process is primarily used here to determine zeropoint offsets between the surveys, and does not correct for the color terms as given in Table 4.  This can be done if there were more \textit{HST} standards that were red ($g-i>1.0$).  Red standards are particularly helpful for the PS1, SDSS and SNLS surveys as SN at $z>0.5$ have $g-i>1.0$.

Instead of waiting for more observations of standard stars to be obtained by \textit{HST}, we can apply the same Supercal process discussed above to catalogs of stars observed by \textit{HST}. Photometry from \textit{HST} is ideal to use for cross-calibration as it is most closely tied to the \textit{HST} Calspec AB system, it is photometrically stable and closely monitored.  The difficulty is that observations taken with \textit{HST} typically are done as part of small programs that have very different objectives, saturation limits and selection effects. In order to bypass the various differences in the small \textit{HST} programs, we could focus on one very large calibrated field: COSMOS \citep{Scoville07}.  Only imaging in a single band, ACS/F814W (roughly $i$), was obtained for the entire field and while the relative calibration across the field is better than 3mmag, there was no effort to define the absolute flux of the observations.  This is not the case for other observations taken of the COSMOS field, so we can use other observations acquired from the Hubble Legacy Archive to determine the \textit{HST} zeropoint of the entire field.  The expected zeropoint does not differ across the sky and any change of the zeropoint with time is negligible.  We find a zeropoint when calibrating to the AB system of $0.268\pm0.003$ mag such that $\textrm{F814W}_{\rm AB}=\textrm{F814W}_{\rm COSMOS}-0.268$ mag.  Applying the Supercal process, we find the calibration offset relative to PS1 of $9\pm10$ mmag with the $i$ band comparison and $6\pm14$ mmag in the $z$ band comparison.  The errors are relatively large because the F814W filter has a relatively different wavelength range and mean effective wavelength than that of either $\ips$ or $\zps$.  More work should be done to tie HST observations to PS1 so it can be included in the Supercal process.

A final question remains about what is the best way to use the Supercal corrections in future cosmological analyses which use SN\,Ia.  A simplistic approach could be not to correct for any of the Supercal offsets, but instead to use them as a systematic uncertainty which one could propagate into the full systematic covariance matrix.  The inherent assumption of this decision is that the current calibration of each survey is optimal, given some uncertainty.  This assumption certainly appears to be false.  Instead, we prefer to correct for the Supercal offsets based on a weighted calibration of the multiple surveys, as done for the average case in Table 4.  This approach creates a homogeneous photometric calibration between the different systems and best ties all the samples to the AB system.
For samples that cannot be included in the Supercal analysis, due to insufficient numbers of stars or due to the lack of a public release of local standards, we think these samples should no longer be included in a cosmological analysis.  
\section{Conclusions}
\label{sec:conclus}

In this paper, we have presented a new method for SN calibration.  We have determined the relative SN-zeropoint offsets between different filters of different systems and show their impact on the measurement of cosmological parameters.  We find that there may be systematic discrepancies between the zeropoints of $1-2\%$ which propagate to up to $5\%$ systematic errors in SN distance.  This systematic error in distance can then result in an average offset of $2-3\%$ in $w$.  The primary discrepancies found are in the $B$ filters of the low-$z$ systems.  More work must be done to better understand these offsets.  The systematic uncertainty of this approach are shown to be of similar magnitude or greater than the statistical uncertainties, and increasing the size and color range of the spectral libraries would be one of the largest improvements to the Supercal process.  Future surveys that are able to more precisely and accurately tie their calibration to the \textit{HST} Calspec standards can use Supercal to combine their sample with past samples.

Overall, the size of the systematic uncertainty on $w$ due to changes in the relative calibration of all surveys used is encouraging for future cosmology studies with SN\,Ia.    While most recent SN\,Ia cosmology studies find that their systematic uncertainties are dominated by issues with calibration, this Supercal analysis shows that the current state may not continue for much longer.


\acknowledgments

We dedicate this paper to Talia Sam Scolnic.  We thank Marc Betoule, Nicolas Regnault, Malcolm Hicken, Andrew Pickles and Mark Phillips for their comments.  This work was performed in part at the Aspen Center for Physics, which is supported by National Science Foundation grant PHY-1066293.

The PS1 Surveys have been made possible through
contributions of the Institute for Astronomy, the University of
Hawaii, the Pan-STARRS Project Office, the Max-Planck Society and its
participating institutes, the Max Planck Institute for Astronomy,
Heidelberg and the Max Planck Institute for Extraterrestrial Physics,
Garching, The Johns Hopkins University, Durham University, the
University of Edinburgh, Queen's University Belfast, the
Harvard-Smithsonian Center for Astrophysics, the Las Cumbres
Observatory Global Telescope Network Incorporated, the National
Central University of Taiwan, the Space Telescope Science Institute,
the National Aeronautics and Space Administration under Grant
No. NNX08AR22G issued through the Planetary Science Division of the
NASA Science Mission Directorate, the National Science Foundation
under Grant No. AST-1238877, the University of Maryland, and Eotvos
Lorand University (ELTE).

R.J.F.\ gratefully acknowledges support from NASA grant 14-WPS14-0048, NSF grant AST-1518052, and the Alfred P.\ Sloan Foundation.

\bibliography{research2}

\begin{thebibliography}{42}
\expandafter\ifx\csname natexlab\endcsname\relax\def\natexlab#1{#1}\fi

\bibitem[{{Betoule} {et~al.}(2014){Betoule}, {Kessler}, {Guy}, {Mosher},
  {Hardin}, {Biswas}, {Astier}, {El-Hage}, {Konig}, {Kuhlmann}, {Marriner},
  {Pain}, {Regnault}, {Balland}, {Bassett}, {Brown}, {Campbell}, {Carlberg},
  {Cellier-Holzem}, {Cinabro}, {Conley}, {D'Andrea}, {DePoy}, {Doi}, {Ellis},
  {Fabbro}, {Filippenko}, {Foley}, {Frieman}, {Fouchez}, {Galbany}, {Goobar},
  {Gupta}, {Hill}, {Hlozek}, {Hogan}, {Hook}, {Howell}, {Jha}, {Le Guillou},
  {Leloudas}, {Lidman}, {Marshall}, {M{\"o}ller}, {Mour{\~a}o}, {Neveu},
  {Nichol}, {Olmstead}, {Palanque-Delabrouille}, {Perlmutter}, {Prieto},
  {Pritchet}, {Richmond}, {Riess}, {Ruhlmann-Kleider}, {Sako}, {Schahmaneche},
  {Schneider}, {Smith}, {Sollerman}, {Sullivan}, {Walton}, \&
  {Wheeler}}]{Betoule14}
{Betoule}, M., {et~al.} 2014, \aap, 568, A22

\bibitem[{{Betoule} {et~al.}(2013){Betoule}, {Marriner}, {Regnault},
  {Cuillandre}, {Astier}, {Guy}, {Balland}, {El Hage}, {Hardin}, {Kessler}, {Le
  Guillou}, {Mosher}, {Pain}, {Rocci}, {Sako}, \& {Schahmaneche}}]{Betoule12}
---. 2013, \aap, 552, A124

\bibitem[{{Bohlin}(1996)}]{1996AJ....111.1743B}
{Bohlin}, R.~C. 1996, \aj, 111, 1743

\bibitem[{{Bohlin} \& {Landolt}(2015)}]{Bohlin15}
{Bohlin}, R.~C., \& {Landolt}, A.~U. 2015, \aj, 149, 122

\bibitem[{{Boulade} {et~al.}(2003){Boulade}, {Charlot}, {Abbon}, {Aune},
  {Borgeaud}, {Carton}, {Carty}, {Da Costa}, {Deschamps}, {Desforge},
  {Eppell{\'e}}, {Gallais}, {Gosset}, {Granelli}, {Gros}, {de Kat}, {Loiseau},
  {Ritou}, {Rouss{\'e}}, {Starzynski}, {Vignal}, \& {Vigroux}}]{Boulade03}
{Boulade}, O., {et~al.} 2003, in Society of Photo-Optical Instrumentation
  Engineers (SPIE) Conference Series, Vol. 4841, Instrument Design and
  Performance for Optical/Infrared Ground-based Telescopes, ed. M.~{Iye} \&
  A.~F.~M. {Moorwood}, 72--81

\bibitem[{{Contreras} {et~al.}(2010){Contreras}, {Hamuy}, {Phillips},
  {Folatelli}, {Suntzeff}, {Persson}, {Stritzinger}, {Boldt}, {Gonz{\'a}lez},
  {Krzeminski}, {Morrell}, {Roth}, {Salgado}, {Jos{\'e} Maureira}, {Burns},
  {Freedman}, {Madore}, {Murphy}, {Wyatt}, {Li}, \& {Filippenko}}]{Contreras10}
{Contreras}, C., {et~al.} 2010, \aj, 139, 519

\bibitem[{{Doi} {et~al.}(2010){Doi}, {Tanaka}, {Fukugita}, {Gunn}, {Yasuda},
  {Ivezi{\'c}}, {Brinkmann}, {de Haars}, {Kleinman}, {Krzesinski}, \& {French
  Leger}}]{Doi10}
{Doi}, M., {et~al.} 2010, \aj, 139, 1628

\bibitem[{{Fukugita} {et~al.}(1996){Fukugita}, {Ichikawa}, {Gunn}, {Doi},
  {Shimasaku}, \& {Schneider}}]{Fukugita96}
{Fukugita}, M., {Ichikawa}, T., {Gunn}, J.~E., {Doi}, M., {Shimasaku}, K., \&
  {Schneider}, D.~P. 1996, \aj, 111, 1748

\bibitem[{{Gunn} {et~al.}(1998){Gunn}, {Carr}, {Rockosi}, {Sekiguchi}, {Berry},
  {Elms}, {de Haas}, {Ivezi{\'c}}, {Knapp}, {Lupton}, {Pauls}, {Simcoe},
  {Hirsch}, {Sanford}, {Wang}, {York}, {Harris}, {Annis}, {Bartozek},
  {Boroski}, {Bakken}, {Haldeman}, {Kent}, {Holm}, {Holmgren}, {Petravick},
  {Prosapio}, {Rechenmacher}, {Doi}, {Fukugita}, {Shimasaku}, {Okada}, {Hull},
  {Siegmund}, {Mannery}, {Blouke}, {Heidtman}, {Schneider}, {Lucinio}, \&
  {Brinkman}}]{Gunn98}
{Gunn}, J.~E., {et~al.} 1998, \aj, 116, 3040

\bibitem[{{Gunn} {et~al.}(2006){Gunn}, {Siegmund}, {Mannery}, {Owen}, {Hull},
  {Leger}, {Carey}, {Knapp}, {York}, {Boroski}, {Kent}, {Lupton}, {Rockosi},
  {Evans}, {Waddell}, {Anderson}, {Annis}, {Barentine}, {Bartoszek}, {Bastian},
  {Bracker}, {Brewington}, {Briegel}, {Brinkmann}, {Brown}, {Carr},
  {Czarapata}, {Drennan}, {Dombeck}, {Federwitz}, {Gillespie}, {Gonzales},
  {Hansen}, {Harvanek}, {Hayes}, {Jordan}, {Kinney}, {Klaene}, {Kleinman},
  {Kron}, {Kresinski}, {Lee}, {Limmongkol}, {Lindenmeyer}, {Long}, {Loomis},
  {McGehee}, {Mantsch}, {Neilsen}, {Neswold}, {Newman}, {Nitta}, {Peoples},
  {Pier}, {Prieto}, {Prosapio}, {Rivetta}, {Schneider}, {Snedden}, \&
  {Wang}}]{Gunn06}
---. 2006, \aj, 131, 2332

\bibitem[{{Gunn} \& {Stryker}(1983)}]{Gunn83}
{Gunn}, J.~E., \& {Stryker}, L.~L. 1983, \apjs, 52, 121

\bibitem[{{Hamuy} {et~al.}(1993){Hamuy}, {Maza}, {Phillips}, {Suntzeff},
  {Wischnjewsky}, {Smith}, {Antezana}, {Wells}, {Gonzalez}, {Gigoux},
  {Navarrete}, {Barrientos}, {Lamontagne}, {della Valle}, {Elias}, {Phillips},
  {Odewahn}, {Baldwin}, {Walker}, {Williams}, {Sturch}, {Baganoff}, {Chaboyer},
  {Schommer}, {Tirado}, {Hernandez}, {Ugarte}, {Guhathakurta}, {Howell},
  {Szkody}, {Schmidtke}, \& {Roth}}]{CT}
{Hamuy}, M., {et~al.} 1993, \aj, 106, 2392

\bibitem[{{Heap} \& {Lindler}(2007)}]{Heap07}
{Heap}, S.~R., \& {Lindler}, D.~J. 2007, in Astronomical Society of the Pacific
  Conference Series, Vol. 374, From Stars to Galaxies: Building the Pieces to
  Build Up the Universe, ed. A.~{Vallenari}, R.~{Tantalo}, L.~{Portinari}, \&
  A.~{Moretti}, 409

\bibitem[{{Hicken} {et~al.}(2009){Hicken}, {Challis}, {Jha}, {Kirshner},
  {Matheson}, {Modjaz}, {Rest}, {Wood-Vasey}, {Bakos}, {Barton}, {Berlind},
  {Bragg}, {Brice{\~n}o}, {Brown}, {Caldwell}, {Calkins}, {Cho}, {Ciupik},
  {Contreras}, {Dendy}, {Dosaj}, {Durham}, {Eriksen}, {Esquerdo}, {Everett},
  {Falco}, {Fernandez}, {Gaba}, {Garnavich}, {Graves}, {Green}, {Groner},
  {Hergenrother}, {Holman}, {Hradecky}, {Huchra}, {Hutchison}, {Jerius},
  {Jordan}, {Kilgard}, {Krauss}, {Luhman}, {Macri}, {Marrone}, {McDowell},
  {McIntosh}, {McNamara}, {Megeath}, {Mochejska}, {Munoz}, {Muzerolle},
  {Naranjo}, {Narayan}, {Pahre}, {Peters}, {Peterson}, {Rines}, {Ripman},
  {Roussanova}, {Schild}, {Sicilia-Aguilar}, {Sokoloski}, {Smalley}, {Smith},
  {Spahr}, {Stanek}, {Barmby}, {Blondin}, {Stubbs}, {Szentgyorgyi}, {Torres},
  {Vaz}, {Vikhlinin}, {Wang}, {Westover}, {Woods}, \& {Zhao}}]{Hicken09a}
{Hicken}, M., {et~al.} 2009, \apj, 700, 331

\bibitem[{{Hicken} {et~al.}(2012){Hicken}, {Challis}, {Kirshner}, {Rest},
  {Cramer}, {Wood-Vasey}, {Bakos}, {Berlind}, {Brown}, {Caldwell}, {Calkins},
  {Currie}, {de Kleer}, {Esquerdo}, {Everett}, {Falco}, {Fernandez},
  {Friedman}, {Groner}, {Hartman}, {Holman}, {Hutchins}, {Keys}, {Kipping},
  {Latham}, {Marion}, {Narayan}, {Pahre}, {Pal}, {Peters}, {Perumpilly},
  {Ripman}, {Sipocz}, {Szentgyorgyi}, {Tang}, {Torres}, {Vaz}, {Wolk}, \&
  {Zezas}}]{Hicken12}
---. 2012, \apjs, 200, 12

\bibitem[{{Holtzman} {et~al.}(2008){Holtzman}, {Marriner}, {Kessler}, {Sako},
  {Dilday}, {Frieman}, {Schneider}, {Bassett}, {Becker}, {Cinabro}, {DeJongh},
  {Depoy}, {Doi}, {Garnavich}, {Hogan}, {Jha}, {Konishi}, {Lampeitl},
  {Marshall}, {McGinnis}, {Miknaitis}, {Nichol}, {Prieto}, {Riess}, {Richmond},
  {Romani}, {Smith}, {Takanashi}, {Tokita}, {van der Heyden}, {Yasuda}, \&
  {Zheng}}]{Holtzman08}
{Holtzman}, J.~A., {et~al.} 2008, \aj, 136, 2306

\bibitem[{{Jha} {et~al.}(2006){Jha}, {Kirshner}, {Challis}, {Garnavich},
  {Matheson}, {Soderberg}, {Graves}, {Hicken}, {Alves}, {Arce}, {Balog},
  {Barmby}, {Barton}, {Berlind}, {Bragg}, {Brice{\~n}o}, {Brown}, {Buckley},
  {Caldwell}, {Calkins}, {Carter}, {Concannon}, {Donnelly}, {Eriksen},
  {Fabricant}, {Falco}, {Fiore}, {Garcia}, {G{\'o}mez}, {Grogin}, {Groner},
  {Groot}, {Haisch}, {Hartmann}, {Hergenrother}, {Holman}, {Huchra},
  {Jayawardhana}, {Jerius}, {Kannappan}, {Kim}, {Kleyna}, {Kochanek},
  {Koranyi}, {Krockenberger}, {Lada}, {Luhman}, {Luu}, {Macri}, {Mader},
  {Mahdavi}, {Marengo}, {Marsden}, {McLeod}, {McNamara}, {Megeath}, {Moraru},
  {Mossman}, {Muench}, {Mu{\~n}oz}, {Muzerolle}, {Naranjo}, {Nelson-Patel},
  {Pahre}, {Patten}, {Peters}, {Peters}, {Raymond}, {Rines}, {Schild},
  {Sobczak}, {Spahr}, {Stauffer}, {Stefanik}, {Szentgyorgyi}, {Tollestrup},
  {V{\"a}is{\"a}nen}, {Vikhlinin}, {Wang}, {Willner}, {Wolk}, {Zajac}, {Zhao},
  \& {Stanek}}]{Jha06}
{Jha}, S., {et~al.} 2006, \aj, 131, 527

\bibitem[{{Kessler} {et~al.}(2009){Kessler}, {Bernstein}, {Cinabro}, {Dilday},
  {Frieman}, {Jha}, {Kuhlmann}, {Miknaitis}, {Sako}, {Taylor}, \&
  {Vanderplas}}]{SNANA}
{Kessler}, R., {et~al.} 2009, \pasp, 121, 1028

\bibitem[{{Landolt}(1992)}]{Landolt92}
{Landolt}, A.~U. 1992, \aj, 104, 340

\bibitem[{{Landolt} \& {Uomoto}(2007)}]{Landolt07}
{Landolt}, A.~U., \& {Uomoto}, A.~K. 2007, \aj, 133, 768

\bibitem[{{Mosher} {et~al.}(2012){Mosher}, {Sako}, {Corlies}, {Folatelli},
  {Frieman}, {Holtzman}, {Jha}, {Kessler}, {Marriner}, {Phillips},
  {Stritzinger}, {Morrell}, \& {Schneider}}]{Mosher12}
{Mosher}, J., {et~al.} 2012, \aj, 144, 17

\bibitem[{{Oke} \& {Gunn}(1983)}]{Oke83}
{Oke}, J.~B., \& {Gunn}, J.~E. 1983, \apj, 266, 713

\bibitem[{{Padmanabhan} {et~al.}(2008){Padmanabhan}, {Schlegel}, {Finkbeiner},
  {Barentine}, {Blanton}, {Brewington}, {Gunn}, {Harvanek}, {Hogg},
  {Ivezi{\'c}}, {Johnston}, {Kent}, {Kleinman}, {Knapp}, {Krzesinski}, {Long},
  {Neilsen}, {Nitta}, {Loomis}, {Lupton}, {Roweis}, {Snedden}, {Strauss}, \&
  {Tucker}}]{Padmanabhan}
{Padmanabhan}, N., {et~al.} 2008, \apj, 674, 1217

\bibitem[{{Perlmutter} {et~al.}(1999){Perlmutter}, {Aldering}, {Goldhaber},
  {Knop}, {Nugent}, {Castro}, {Deustua}, {Fabbro}, {Goobar}, {Groom}, {Hook},
  {Kim}, {Kim}, {Lee}, {Nunes}, {Pain}, {Pennypacker}, {Quimby}, {Lidman},
  {Ellis}, {Irwin}, {McMahon}, {Ruiz-Lapuente}, {Walton}, {Schaefer}, {Boyle},
  {Filippenko}, {Matheson}, {Fruchter}, {Panagia}, {Newberg}, {Couch}, \&
  {Project}}]{Perlmutter99}
{Perlmutter}, S., {et~al.} 1999, \apj, 517, 565

\bibitem[{{Pickles}(1998)}]{Pickles98}
{Pickles}, A.~J. 1998, \pasp, 110, 863

\bibitem[{{Planck Collaboration} {et~al.}(2014){Planck Collaboration}, {Ade},
  {Aghanim}, {Armitage-Caplan}, {Arnaud}, {Ashdown}, {Atrio-Barandela},
  {Aumont}, {Baccigalupi}, {Banday}, \& et~al.}]{Planck14}
{Planck Collaboration}, {et~al.} 2014, \aap, 571, A16

\bibitem[{{Regnault} {et~al.}(2009){Regnault}, {Conley}, {Guy}, {Sullivan},
  {Cuillandre}, {Astier}, {Balland}, {Basa}, {Carlberg}, {Fouchez}, {Hardin},
  {Hook}, {Howell}, {Pain}, {Perrett}, \& {Pritchet}}]{Regnault09}
{Regnault}, N., {et~al.} 2009, \aap, 506, 999

\bibitem[{{Rest} {et~al.}(2014){Rest}, {Scolnic}, {Foley}, {Huber}, {Chornock},
  {Narayan}, {Tonry}, {Berger}, {Soderberg}, {Stubbs}, {Riess}, {Kirshner},
  {Smartt}, {Schlafly}, {Rodney}, {Botticella}, {Brout}, {Challis}, {Czekala},
  {Drout}, {Hudson}, {Kotak}, {Leibler}, {Lunnan}, {Marion}, {McCrum},
  {Milisavljevic}, {Pastorello}, {Sanders}, {Smith}, {Stafford}, {Thilker},
  {Valenti}, {Wood-Vasey}, {Zheng}, {Burgett}, {Chambers}, {Denneau}, {Draper},
  {Flewelling}, {Hodapp}, {Kaiser}, {Kudritzki}, {Magnier}, {Metcalfe},
  {Price}, {Sweeney}, {Wainscoat}, \& {Waters}}]{Rest14}
{Rest}, A., {et~al.} 2014, \apj, 795, 44

\bibitem[{{Riess} {et~al.}(1998){Riess}, {Filippenko}, {Challis},
  {Clocchiatti}, {Diercks}, {Garnavich}, {Gilliland}, {Hogan}, {Jha},
  {Kirshner}, {Leibundgut}, {Phillips}, {Reiss}, {Schmidt}, {Schommer},
  {Smith}, {Spyromilio}, {Stubbs}, {Suntzeff}, \& {Tonry}}]{Riess98}
{Riess}, A.~G., {et~al.} 1998, \aj, 116, 1009

\bibitem[{{Riess} {et~al.}(1999){Riess}, {Kirshner}, {Schmidt}, {Jha},
  {Challis}, {Garnavich}, {Esin}, {Carpenter}, {Grashius}, {Schild}, {Berlind},
  {Huchra}, {Prosser}, {Falco}, {Benson}, {Brice{\~n}o}, {Brown}, {Caldwell},
  {dell'Antonio}, {Filippenko}, {Goodman}, {Grogin}, {Groner}, {Hughes},
  {Green}, {Jansen}, {Kleyna}, {Luu}, {Macri}, {McLeod}, {McLeod}, {McNamara},
  {McLean}, {Milone}, {Mohr}, {Moraru}, {Peng}, {Peters}, {Prestwich},
  {Stanek}, {Szentgyorgyi}, \& {Zhao}}]{Riess99}
---. 1999, \aj, 117, 707

\bibitem[{{Schlafly} \& {Finkbeiner}(2011)}]{Schlafly11}
{Schlafly}, E.~F., \& {Finkbeiner}, D.~P. 2011, \apj, 737, 103

\bibitem[{{Schlafly} {et~al.}(2012){Schlafly}, {Finkbeiner}, {Juri{\'c}},
  {Magnier}, {Burgett}, {Chambers}, {Grav}, {Hodapp}, {Kaiser}, {Kudritzki},
  {Martin}, {Morgan}, {Price}, {Rix}, {Stubbs}, {Tonry}, \&
  {Wainscoat}}]{Schlafly12}
{Schlafly}, E.~F., {et~al.} 2012, \apj, 756, 158

\bibitem[{{Scolnic} {et~al.}(2014{\natexlab{a}}){Scolnic}, {Rest}, {Riess},
  {Huber}, {Foley}, {Brout}, {Chornock}, {Narayan}, {Tonry}, {Berger},
  {Soderberg}, {Stubbs}, {Kirshner}, {Rodney}, {Smartt}, {Schlafly},
  {Botticella}, {Challis}, {Czekala}, {Drout}, {Hudson}, {Kotak}, {Leibler},
  {Lunnan}, {Marion}, {McCrum}, {Milisavljevic}, {Pastorello}, {Sanders},
  {Smith}, {Stafford}, {Thilker}, {Valenti}, {Wood-Vasey}, {Zheng}, {Burgett},
  {Chambers}, {Denneau}, {Draper}, {Flewelling}, {Hodapp}, {Kaiser},
  {Kudritzki}, {Magnier}, {Metcalfe}, {Price}, {Sweeney}, {Wainscoat}, \&
  {Waters}}]{Scolnic14b}
{Scolnic}, D., {et~al.} 2014{\natexlab{a}}, \apj, 795, 45

\bibitem[{{Scolnic} {et~al.}(2014{\natexlab{b}}){Scolnic}, {Riess}, {Foley},
  {Rest}, {Rodney}, {Brout}, \& {Jones}}]{S14a}
{Scolnic}, D.~M., {Riess}, A.~G., {Foley}, R.~J., {Rest}, A., {Rodney}, S.~A.,
  {Brout}, D.~J., \& {Jones}, D.~O. 2014{\natexlab{b}}, \apj, 780, 37

\bibitem[{{Scoville} {et~al.}(2007){Scoville}, {Aussel}, {Brusa}, {Capak},
  {Carollo}, {Elvis}, {Giavalisco}, {Guzzo}, {Hasinger}, {Impey}, {Kneib},
  {LeFevre}, {Lilly}, {Mobasher}, {Renzini}, {Rich}, {Sanders}, {Schinnerer},
  {Schminovich}, {Shopbell}, {Taniguchi}, \& {Tyson}}]{Scoville07}
{Scoville}, N., {et~al.} 2007, \apjs, 172, 1

\bibitem[{{Smith} {et~al.}(2002){Smith}, {Tucker}, {Kent}, {Richmond},
  {Fukugita}, {Ichikawa}, {Ichikawa}, {Jorgensen}, {Uomoto}, {Gunn}, {Hamabe},
  {Watanabe}, {Tolea}, {Henden}, {Annis}, {Pier}, {McKay}, {Brinkmann}, {Chen},
  {Holtzman}, {Shimasaku}, \& {York}}]{Smith02}
{Smith}, J.~A., {et~al.} 2002, \aj, 123, 2121

\bibitem[{{Stritzinger} {et~al.}(2011){Stritzinger}, {Phillips}, {Boldt},
  {Burns}, {Campillay}, {Contreras}, {Gonzalez}, {Folatelli}, {Morrell},
  {Krzeminski}, {Roth}, {Salgado}, {DePoy}, {Hamuy}, {Freedman}, {Madore},
  {Marshall}, {Persson}, {Rheault}, {Suntzeff}, {Villanueva}, {Li}, \&
  {Filippenko}}]{Stritzinger11}
{Stritzinger}, M.~D., {et~al.} 2011, \aj, 142, 156

\bibitem[{{Stubbs} {et~al.}(2010){Stubbs}, {Doherty}, {Cramer}, {Narayan},
  {Brown}, {Lykke}, {Woodward}, \& {Tonry}}]{Stubbs10}
{Stubbs}, C.~W., {Doherty}, P., {Cramer}, C., {Narayan}, G., {Brown}, Y.~J.,
  {Lykke}, K.~R., {Woodward}, J.~T., \& {Tonry}, J.~L. 2010, \apjs, 191, 376

\bibitem[{{Stubbs} \& {Tonry}(2006)}]{Stubbs06}
{Stubbs}, C.~W., \& {Tonry}, J.~L. 2006, \apj, 646, 1436

\bibitem[{{Tonry} {et~al.}(2012{\natexlab{a}}){Tonry}, {Stubbs}, {Kilic},
  {Flewelling}, {Deacon}, {Chornock}, {Berger}, {Burgett}, {Chambers},
  {Kaiser}, {Kudritzki}, {Hodapp}, {Magnier}, {Morgan}, {Price}, \&
  {Wainscoat}}]{Tonry11}
{Tonry}, J.~L., {et~al.} 2012{\natexlab{a}}, \apj, 745, 42

\bibitem[{{Tonry} {et~al.}(2012{\natexlab{b}}){Tonry}, {Stubbs}, {Lykke},
  {Doherty}, {Shivvers}, {Burgett}, {Chambers}, {Hodapp}, {Kaiser},
  {Kudritzki}, {Magnier}, {Morgan}, {Price}, \& {Wainscoat}}]{Tonry12}
---. 2012{\natexlab{b}}, \apj, 750, 99

\bibitem[{{Tucker} {et~al.}(2006){Tucker}, {Kent}, {Richmond}, {Annis},
  {Smith}, {Allam}, {Rodgers}, {Stute}, {Adelman-McCarthy}, {Brinkmann}, {Doi},
  {Finkbeiner}, {Fukugita}, {Goldston}, {Greenway}, {Gunn}, {Hendry}, {Hogg},
  {Ichikawa}, {Ivezi{\'c}}, {Knapp}, {Lampeitl}, {Lee}, {Lin}, {McKay},
  {Merrelli}, {Munn}, {Neilsen}, {Newberg}, {Richards}, {Schlegel},
  {Stoughton}, {Uomoto}, \& {Yanny}}]{Tucker06}
{Tucker}, D.~L., {et~al.} 2006, Astronomische Nachrichten, 327, 821

\end{thebibliography}

\end{document}